\documentclass[aps,preprintnumbers,superscriptaddress,runinaddress,amsmath,amssymb,showpacs,floatfix,twocolumn]{revtex4-1}

\usepackage{graphicx}
\usepackage{amsmath}

\newcommand{\dt}{dt}

\begin{document}

\title{Lattice baryon spectroscopy with multi-particle interpolators}


\author{Adrian L. Kiratidis}
\email{adrian.kiratidis@adelaide.edu.au} 
\author{Waseem Kamleh} 
\author{Derek B. Leinweber} 
\author{Benjamin J. Owen}
  \affiliation{Special Research Centre for the Subatomic Structure of
  Matter, Department of Physics, University of Adelaide,
  South Australia 5005, Australia.}

\begin{abstract}
In $2 + 1$ flavour lattice QCD the spectrum of the nucleon is
presented for both parities using local meson-baryon type
interpolating fields in addition to the standard three-quark nucleon
interpolators.	
The role of local five-quark operators in extracting the nucleon
excited state spectrum via correlation matrix techniques is explored
on dynamical gauge fields with $m_{\pi} = 293\textrm{ MeV},$ leading
to the observation of a state in the region of the non-interacting S-wave $N\pi$
scattering threshold in the negative-parity sector.  Furthermore, the
robustness of the variational technique is examined by studying the
spectrum on a variety of operator bases. Fitting a single-state ansatz
to the eigenstate-projected correlators provides robust energies for
the low-lying spectrum that are essentially invariant despite being
extracted from qualitatively different bases.
\end{abstract}

\pacs{11.15.Ha,12.38.-t,12.38.Gc}

\maketitle

\section{Introduction}

Lattice QCD is currently the only known \textit{ab-initio}
non-perturbative approach to study the fundamental quantum field
theory governing hadron properties, Quantum Chromodynamics (QCD).  While the
ability to obtain ground state masses is well-understood, an
accurate extraction of excited states and multi-particle thresholds
remains a challenge.

The use of variational techniques~\cite{Michael:1985ne, Luscher:1990ck}
to study the nucleon excited state spectrum has seen remarkable
success in recent years. The key feature of these techniques is to
begin with a basis of different operators that couple to the quantum
numbers of a given state, and then construct different linear
combinations of these operators in order to isolate the ground and higher
excited states in that channel.

The positive-parity nucleon channel has been of significant interest
to the lattice community~\cite{Liu:2014jua, Roberts:2013ipa,
  Bauer:2012at, Edwards:2011jj, Roberts:2011ym, Mahbub:2010rm}.  In
particular the first positive-parity $J^P = {\frac{1}{2}}^{+}$
excitation of the nucleon, known as the Roper resonance $N^{*}(1440)$,
remains a puzzle.  In constituent quark models the Roper resonance
lies above the lowest-lying negative-parity state~\cite{Isgur:1977ef,
  Isgur:1978wd, Glozman:1995fu}, the $N^{*}(1535)$, whereas in Nature
it lies $95\textrm{ MeV}$ below the resonant state.  This has led to
speculation about the true nature of this state, with suggestions it is
a baryon with explicitly excited gluon fields, or that it can be
understood with meson-baryon dynamics via a meson-exchange
model~\cite{Speth:2000zf}.

In simple quark models, the Roper is identified with an $N=2$ radial
excitation of the nucleon. Within the variational technique, the
choice of an appropriate operator basis is critical to obtaining the
complete spectrum of low-lying excited states. Recall that we can
expand any radial function using a basis of Gaussians of different
widths $f(|\vec{r}|) = \sum_i c_i e^{-\varepsilon_i r^2}.$ This leads
to the use of Gaussian-smeared fermion sources with a variety of
widths~\cite{Burch:2004he}, providing an operator basis that is highly
suited to accessing radial excitations. The CSSM lattice collaboration
has used this technique to study the nucleon excited state spectrum~\cite{Mahbub:2010jz,Mahbub:2012ri}. In particular,
the CSSM studies were the first to demonstrate that the inclusion of
very wide quark fields (formed with large amounts of Gaussian
smearing) is critical to isolating the first positive-parity nucleon
excited state~\cite{Mahbub:2009aa,Mahbub:2010rm}.  This state was shown to
have a quark probability distribution consistent with an $N=2$ radial
excitation in Ref.~\cite{Roberts:2013oea}. This work also examined the
quark probability distributions for higher positive-parity nucleon
excited states, revealing that the combination of Gaussian sources of
different widths allows for the formation of the nodal structures that
characterise the different radial excitations.

The negative-parity nucleon channel with its two low-lying resonances,
the $N^{*}(1535)$ and $N^{*}(1650),$ has also been of
significant interest~\cite{Bruns:2010sv, Edwards:2011jj, Lang:2012db,
  Mahbub:2012ri, Mahbub:2013bba}.  These $S_{11}$ states are in
agreement with $SU(6)$ based quark model predictions, making an $\textit{ab-initio}$ study of the low-lying negative-parity spectrum a
potentially rewarding endeavour. Importantly, at near physical quark
masses the non-interacting $\pi{N}$ scattering threshold lies below the lowest lying
negative-parity state, making it a natural place to look for the
presence of multi-particle energy levels in the extracted spectrum.

Until recently, the majority of the work in these channels has been
performed with three-quark interpolating fields, and in the full
quantum field theory these interpolators couple to more exotic
meson-baryon components such as the aforementioned $\pi{N}$ via
sea-quark loop interactions.  However, baryon studies have found that
the couplings of single hadron type operators to hadron-hadron type
components, suppressed by the lattice volume as $1/\sqrt{V},$ are
sufficiently low so as to make it difficult to observe states associated with scattering thresholds ~\cite{Edwards:2011jj,Mahbub:2013bba}.  Moreover, there is a
question as to what extent the presence of multi-particle states might
interfere with the extraction of nearby resonances.

One solution is to explicitly include hadron-hadron type
interpolators~\cite{Morningstar:2013bda, Lang:2012db} by combining
single-hadron operators with the relevant momentum.  This creates an
operator that necessarily has a high overlap with the scattering state
of interest thereby enabling its extraction.  Instead, in this work we
aim to construct meson-baryon type interpolators without explicitly
projecting single-hadron momenta, and investigate the role that the
resulting operator plays in the calculation of the nucleon spectrum.
Using these operators we construct a basis containing both three- and
five-quark operators, and perform spectroscopic calculations utilising
a variety of different sub-bases.  Examining the resulting spectra
then provides an excellent opportunity to both study the role of our
multi-particle operators and test the robustness of the variational
techniques employed.

Following the outline of standard variational analyses in
Section~\ref{sect:CorrelationMatrixTechniques}, we construct these
hadron-hadron type interpolators in the form of five-quark operators
in Section~\ref{sect:MultiParticleStateContributions}.  We then develop
a method for smearing elements of the stochastically estimated loop
propagators at $x$, $S(x,x)$ in
Section~\ref{sect:LoopPropagatorTechniques}.  These necessarily arise
with the introduction of our five-quark interpolating fields, due to
the presence of creation quark fields in our annihilation operator and
vice versa.  Having covered the technology required for a
spectroscopic calculation, we then outline our simulation details in Section~\ref{sect:Simulation Details} and present nucleon spectra for both
parities in Section~\ref{sect:Results}.

\section{Correlation Matrix Techniques}
\label{sect:CorrelationMatrixTechniques}

Correlation matrix techniques~\cite{Michael:1985ne, Luscher:1990ck}
are now well-established as a method for studying the excited state
hadron spectrum. The underlying principle is to begin with a
sufficiently large basis of $N$ operators (so as to span the space of
the states of interest within the spectrum) and construct an $N \times
N$ matrix of cross correlation functions,
\begin{equation}
\label{defn:CM}
\mathcal{G}_{ij}(\vec{p},t) = \sum_{\vec{x}}\textrm{e}^{-i\vec{p}\cdot\vec{x}}\,\big\langle\,\Omega\, \big| \,\chi_{i}(\vec{x},t)\,\overline{\chi}_{j}(\vec{0},t_{src})\, \big| \,\Omega\, \big\rangle.
\end{equation}
After selecting $\vec{p}=\vec{0}$ and projecting to a specific parity 
with the operator 
\begin{equation}\label{ParityProjector}
\Gamma_{\pm} = \frac{1}{2}\,(\gamma_{0} \pm I)\,,
\end{equation}
we can write the correlator as a sum of exponentials,
\begin{equation}
\label{GijSum}
\mathcal{G}_{ij}(t) = \sum_{\alpha}\lambda^{\alpha}_{i}\,\bar{\lambda}^{\alpha}_{j}\,\textrm{e}^{-m_{\alpha}t},
\end{equation}
where $\alpha$ enumerates the energy eigenstates of mass $m_{\alpha}$
and $\bar{\lambda}^{\alpha}_{j}$ and $\lambda^{\alpha}_{i}$ are the
couplings of our creation and annihilation operators
$\overline{\chi}_{j}$ and $\chi_{i}$ at the source and sink
respectively.
We then search for a linear combination of operators
\begin{equation}
\bar{\phi}^{\alpha} = \bar{\chi}_{j}\,u^{\alpha}_{j} \qquad \textrm{ and } \qquad \phi^{\alpha} = \chi_{i}\,v^{\alpha}_{i}
\end{equation}  
such that $\phi$ and $\bar{\phi}$ couple to a single energy eigenstate.  That is, we require
\begin{equation}
\big\langle\,\Omega\, \big| \,\phi^{\alpha}\,\big| \beta \big\rangle \propto \delta^{\alpha\beta}.
\end{equation}
One can then see from Eq. (\ref{GijSum}) that 
\begin{equation}
\mathcal{G}_{ij}(t_{0} + \dt)\,u^{\alpha}_{j} = \textrm{e}^{-m_{\alpha}\dt}\,\mathcal{G}_{ij}(t_{0})\,u^{\alpha}_{j},
\end{equation}
and hence the required values for $u^{\alpha}_{j}$ and
$v^{\alpha}_{i}$ for a given choice of variational parameters
$(t_0,\dt)$ can be obtained by solving the eigenvalue equations
\begin{align}
\label{E-value-eq-L}
\big[\mathcal{G}^{-1}(t_{0})\,\mathcal{G}(t_{0} + \dt)\big]_{ij}\,u^{\alpha}_{j} &= c^{\alpha}\,u^{\alpha}_{i}\\
\label{E-value-eq-R}
v^{\alpha}_{i}\,\big[\mathcal{G}(t_{0} + \dt)\,\mathcal{G}^{-1}(t_{0})\big]_{ij} &=
c^{\alpha}\,v^{\alpha}_{j},
\end{align}
where the eigenvalue is $c^{\alpha} = \textrm{e}^{-m_{\alpha}\dt}$.  In the ensemble average $\mathcal{G}_{ij}$ is a symmetric matrix.  We work with the improved estimator $\frac{1}{2}\,(\mathcal{G}_{ij} + \mathcal{G}_{ji})$ ensuring the eigenvalues of Eqs. (\ref{E-value-eq-L}) and (\ref{E-value-eq-R}) are equal.
As our correlation matrix is diagonalised at $t_{0}$ and $t_{0} +
\dt$ by the eigenvectors $u^{\alpha}_{j}$ and $v^{\alpha}_{i}$
we can obtain the eigenstate-projected correlator as a function of Euclidean time
\begin{equation}
\mathcal{G}^{\alpha}(t) = v^{\alpha}_{i}\,\mathcal{G}_{ij}(t)\,u^{\alpha}_{j},
\end{equation}
which can then be used to extract masses.  Moreover, the analysis can
be performed on a symmetric matrix with orthogonal eigenvectors.  More
details can be found in Ref.~\cite{Mahbub:2012ri}.

\begin{figure}[htp]
  {\includegraphics[width=0.48\textwidth]{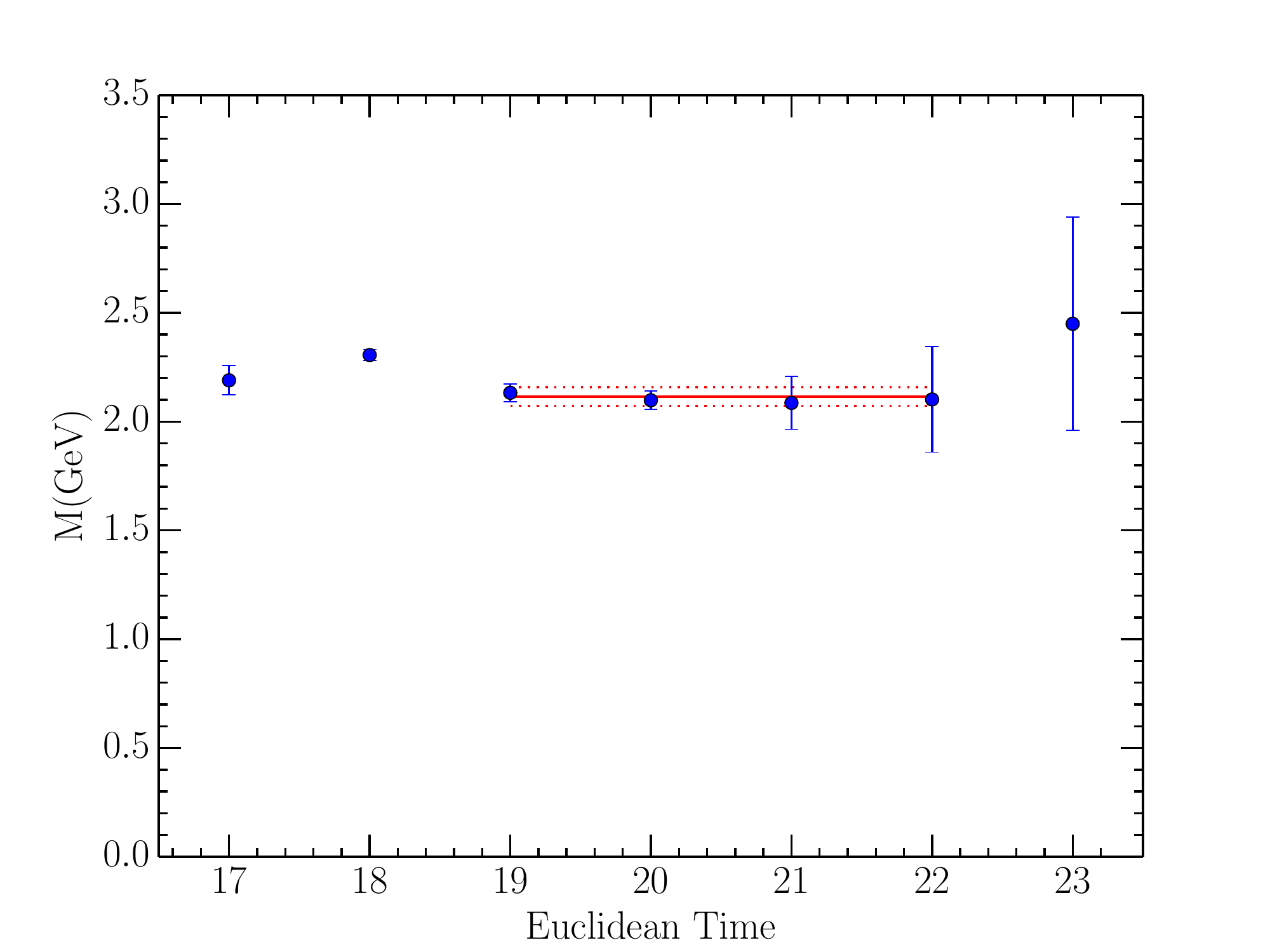}}\\
  {\includegraphics[width=0.48\textwidth]{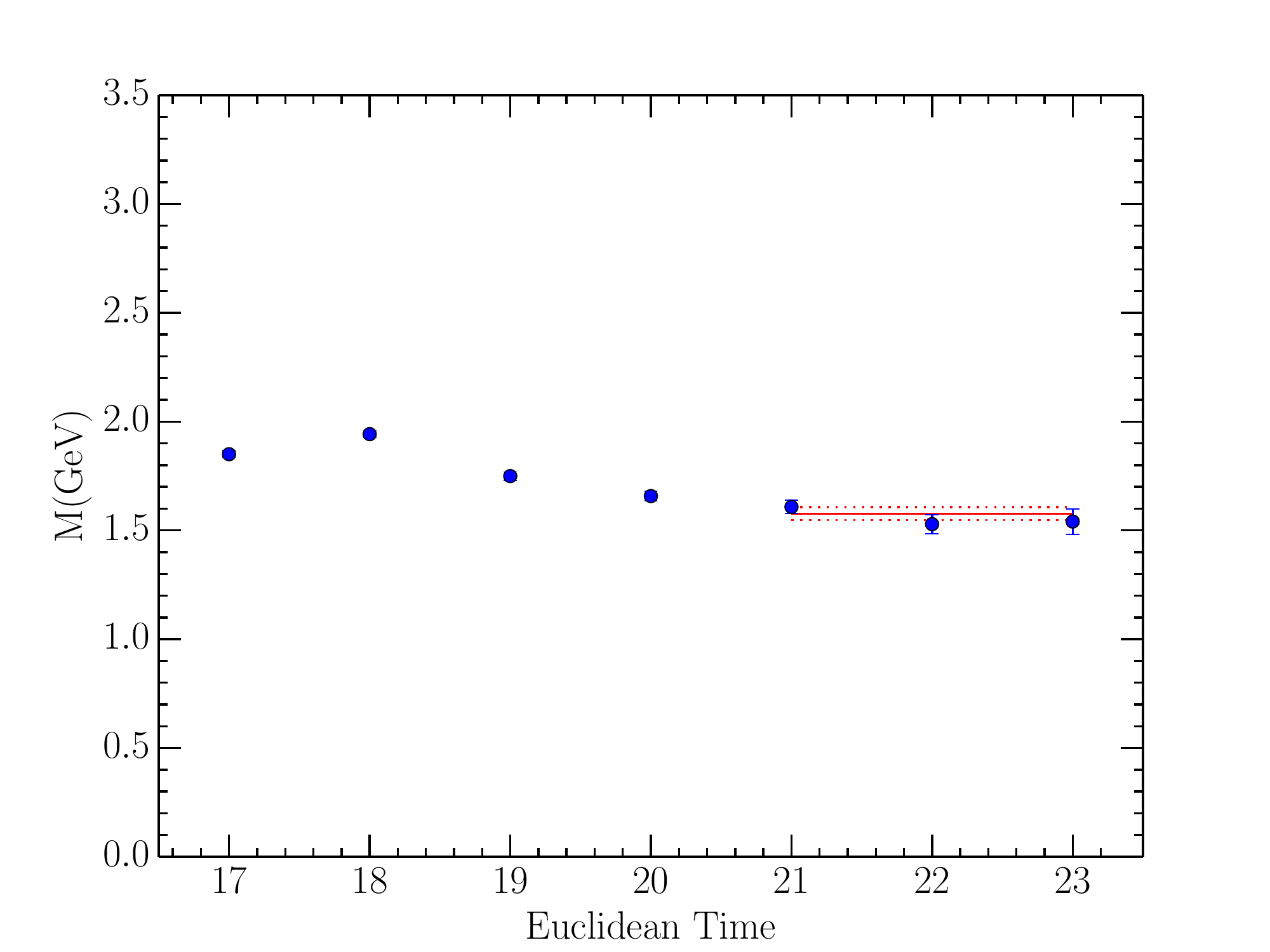}}
 \caption{(Colour online).  Typical effective mass fits for positive-parity (top) and negative-parity (bottom) nucleon excitations.  The top plot shows a fit to the first positive-parity excitation of the $4 \times 4$ correlation matrix obtained from basis 1 described in Table \ref{table:BasisTable} of Section \ref{sect:Simulation Details}.  The fitted mass of 2.11(4) GeV is shown as a green circle for basis 1 in Fig. \ref{fig:Masses+}, and provides $\chi^{2}$/dof $=0.17$.  The bottom plot shows a fit to the lowest-lying state in the negative-parity sector.  It is sourced from the $6 \times 6$ correlation matrix obtained from basis 3 described in Table \ref{table:BasisTable} of Section \ref{sect:Simulation Details}.  The fitted mass of 1.58(3) GeV is shown as a blue square for basis 3 in Fig. \ref{fig:Masses-}, and corresponds to $\chi^{2}$/dof $=0.87$.  Note, an earlier fit including $t = 20$ provides $\chi^{2}$/dof $=1.22$, reflecting the systematic drift in the effective mass at early times.}
  \label{fig:MassFits}
\end{figure}

At this point we note that if the operator basis does
not appropriately span the low-lying spectrum, $\mathcal{G}^{\alpha}(t)$
may contain a mixture of two or more energy eigenstates. There are a
number of scenarios in which this might occur:
\begin{itemize}
\item At early Euclidean times the number of states strongly
  contributing to the correlation matrix may be (much) larger than the
  number of operators in the basis.

\item There may be energy eigenstates present that do not couple or
  only couple weakly to the operators used. In particular, it is well
  known that local three-quark interpolating fields couple poorly to
  multi-hadron scattering states.

\item The nature of the operators selected may be such that it is not
  possible to construct a linear combination with the appropriate
  structure to isolate a particular state.
\end{itemize}
It is important to have a strategy to ensure that one can
accurately obtain eigenstate energies from the correlation matrix.
The method we use is to analyse the effective energies of
different states from the eigenstate-projected correlators,
\begin{equation}
E^\alpha(t) = \frac{1}{n}\log \frac{\mathcal{G}^{\alpha}(t)}{\mathcal{G}^{\alpha}(t+n)}
\end{equation}
which is constant in regions where the correlator is dominated by a
single state.  Neighbouring time slices in the correlation functions are highly correlated in Euclidean time, and require a covariance-matrix based $\chi^2$ analysis.  The best unbiased estimate corresponds to a $\chi^{2}$/dof $\approx$ 1.  We therefore endeavour to obtain a plateau fit of the effective mass with the $\chi^{2}$/dof close to one.  In considering an upper limit for the fit, points with errors bars larger than the central value are discarded.  Fits with $\chi^{2}$/dof $> 1.20$ are rejected, as these fits have significant contamination from nearby states not yet isolated in the correlation matrix analysis~\cite{Mahbub:2013bba}.  We do not enforce a lower bound on acceptable $\chi^{2}$/dof as small values typically reflect large uncertainties rather than an incorrect result associated with a systematic error.  Typically, plateaus commence three or four time slices after the source, near the regime where the generalised eigenvector analysis of the correlation matrix is done.  Figure \ref{fig:MassFits} illustrates typical effective mass fits for positive and negative parity states.  Further details of this method can found in Ref.~\cite{Mahbub:2010lha}.

As we will demonstrate, a careful covariance-matrix based $\chi^2$ analysis to fit the single-state ansatz ensures a robust extraction of the
eigenstate energies.  
The physics underpinning this robustness is elucidated in detail in Section \ref{sect:Results}.

The CSSM lattice collaboration has used this technique in the calculation of the nucleon spectra in both the
positive~\cite{Mahbub:2010rm} and negative-parity
channels~\cite{Mahbub:2012ri} with standard three-quark interpolators.  While largely successful at identifying towers of excited
states that would be associated with resonances in Nature, it has been
shown that with three-quark operators alone it is difficult to detect states near
multi-particle scattering energy levels~\cite{Mahbub:2013bba}.  The
concern is that the operator basis doesn't have sufficient overlap
with meson-baryon type components, highlighting the need for studies
with multi-hadron operators.
%


%
\section{Multi-particle State Contributions}
\label{sect:MultiParticleStateContributions}
In order to further elucidate the situation, we consider a simple two-component toy model which consists of two QCD energy eigenstates, $|\, a \rangle$ and $|\, b \rangle$.  We then suppose that $|\, a \rangle$ and $|\, b \rangle$ are given by
\begin{align}
\big|\, a \big\rangle &= \;\;\, \cos\theta\, \big|\, 1 \big\rangle + \sin\theta\: \big|\, 2 \big\rangle\, , \\
\big|\, b \big\rangle &= -\sin\theta\, \big|\, 1 \big\rangle + \cos\theta\, \big|\, 2 \big\rangle\, , 
\end{align}
where $|\, 1 \rangle$ and $|\, 2 \rangle$ denote a single-hadron and meson-baryon type component respectively, while $\theta$ is some arbitrary mixing angle.
Now imagine performing a spectroscopic calculation with an interpolating field $\chi_{3}$ that only has substantial overlap with $|\, 1 \rangle$.  That is,
\begin{equation}
\big\langle \Omega\, \big |\, \chi_{3} \, \big |\, 1 \rangle \propto C \, \quad \mbox{and}
\qquad 
\big\langle \Omega\, \big |\, \chi_{3} \, \big |\, 2 \rangle \ll C \, ,
\end{equation}
for some constant $C$.  When $\overline{\chi}_{3}$ acts on the vacuum we therefore create a state that is superposition of the true energy eigenstates given by
\begin{equation}
\big|\, 1 \big\rangle = \cos\theta\, \big|\, a \big\rangle - \sin\theta\: \big|\, b \big\rangle \, .
\end{equation}
In the absence of an operator that has substantial overlap with $|\, 2
\rangle$, it becomes impossible to separate out the true QCD
eigenstates of interest.  This naturally leads to two points of
concern.  Firstly, one cannot extract states with a significant $|\, 2
\rangle$ component and secondly there is possibly contamination of the
states that are extracted.  When performing baryon spectroscopy it
therefore becomes desirable to include interpolating fields that we
expect to have substantial overlap with multi-particle meson-baryon
type states~\cite{Lang:2012db}.  While projecting single-hadron
momenta in a multi-hadron operator allows for a clean extraction of states associated with
scattering thresholds, the influence of local five-quark operators
(without explicit momenta assigned to each hadron) on the spectrum
is less intuitive. It is the purpose of this study to examine the role local five-quark operators play in the spectrum, and to thereby
test the robustness of our variational method.

Starting with standard $N$ and $\pi$ interpolators we use the Clebsch-Gordan
coefficients to project isospin $I = 1/2, I_{3} = +1/2$ and write down
the general form of our meson-baryon interpolating
fields~\cite{Kiratidis:2012mr, Kamleh:2014nxa},
\begin{align}\label{Proton5QrkOpInterpolator}
\chi_{N\pi}(x) &= \frac{1}{\sqrt{6}}\,\epsilon^{abc}\,\gamma_{5}\times\nonumber\\
&\quad\bigg\{2\big[u^{Ta}(x)\,\Gamma_{1}\,d^{b}(x)\big]\,\Gamma_{2}\,d^{c}(x)\,\big[\bar{d}^{e}(x)\,\gamma_{5}\,u^{e}(x)\big]\nonumber\\
&\quad - \big[u^{Ta}(x)\,\Gamma_{1}\,d^{b}(x)\big]\,\Gamma_{2}\,u^{c}(x)\,\big[\bar{d}^{e}(x)\,\gamma_{5}\,d^{e}(x)\,\big]\nonumber\\
&\quad + \big[u^{Ta}(x)\,\Gamma_{1}\,d^{b}(x)\big]\,\Gamma_{2}\,u^{c}(x)\,\big[\,\bar{u}(x)^{e}\,\gamma_{5}\,u^{e}(x)\big]\bigg\},
\end{align}
providing us with two five-quark operators, denoted $\chi_{5}$ and
$\chi^{\prime}_{5}$ which correspond to $(\Gamma_{1}, \Gamma_{2}) =
(C\gamma_{5}, \textrm{I})$ and $(\Gamma_{1}, \Gamma_{2}) =
(C,\gamma_{5})$ respectively.  The square brackets around the diquark contraction denote a Dirac scalar.  Under a parity transformation
\begin{equation}
x \rightarrow \tilde{x} = (x_{0}, -\vec{x})\,,
\end{equation}
and the quark fields $\psi(x)$ and $\bar{\psi}(x)$ transform as
\begin{align}
\psi(x) &\rightarrow \mathcal{P}\,\psi(x)\,\mathcal{P}^{\dagger} = \gamma_{0}\,\psi(\tilde{x})\,,\nonumber\\
\bar{\psi}(x) &\rightarrow \mathcal{P}\,\bar{\psi}(x)\,\mathcal{P}^{\dagger} = \bar{\psi}(\tilde{x})\,\gamma_{0}.
\end{align}
Applying a parity transformation to the standard pion interpolator $\chi_{\pi}(x) = \bar{\psi}(x)\gamma_{5}\psi(x)$, and the nucleon interpolators of type $\chi_{N}(x) = \big[\psi^{T}(x)(C\gamma_{5})\psi(x)\big]\psi(x)$ of Eq. (\ref{NucleonInterps}) we find
\begin{align}
\chi_{\pi}(x) &\rightarrow -\bar{\psi}(\tilde{x})\,\gamma_{5}\,\psi(\tilde{x}) = -\chi_{\pi}(\tilde{x})\,,\nonumber\\
\chi_{N}(x) &\rightarrow \big[\psi^{T}(\tilde{x})\,(C\gamma_{5})\,\psi(\tilde{x})\big]\gamma_{0}\,\psi(\tilde{x}) = \gamma_{0}\,\chi_{N}(\tilde{x}).
\end{align}
Thus the pion interpolator transforms negatively under parity.  To ensure our five-quark baryon interpolator formed from the product of pion and nucleon interpolators, transforms in the appropriate manner, the prefactor of $\gamma_{5}$ is included in Eq. (\ref{Proton5QrkOpInterpolator}).
That is, both our three-quark and five-quark nucleon operators have the same parity transformation properties and hence can be combined in a correlation matrix.  This also ensures the standard parity projector of Eq. (\ref{ParityProjector}) applies to our five-quark interpolators.
\begin{figure}[t!!]
  \centering
  {\includegraphics[width=0.44\textwidth]{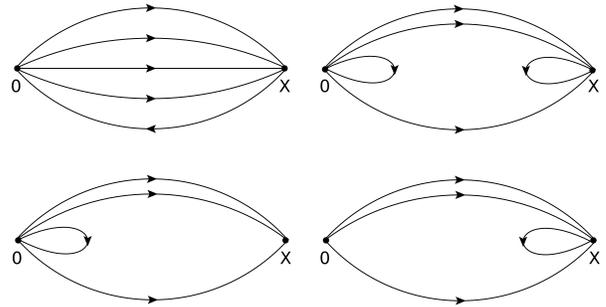}}
 \caption{The Feynman diagrams considered following the introduction of five-quark interpolating fields to standard three-quark operators.}
  \label{fig:FeynmanDiag}
\end{figure}

The presence of creation quark fields in our annihilation
interpolating field and vice versa then leads to the requirement of
calculating the more computationally intense loop propagators, in
order to compute the diagrams in Fig.~\ref{fig:FeynmanDiag}.  The
literature contains different ways of dealing with such diagrams such
as distillation~\cite{Peardon:2009gh}, and various schemes such as the
Laplacian Heaviside (LapH) smearing method~\cite{Morningstar:2011ka}.
Here we will stochastically estimate inverse matrix elements fully
diluting in spin, colour and time as outlined below.

%
%
\section{Loop Propagator Techniques}
\label{sect:LoopPropagatorTechniques}

As observed in the preceding section, spectroscopic calculations that
involve the five-quark operators $\chi_{5}$ and $\chi_{5}^{\prime}$
necessarily involve the determination of loop propagators at $x,$
denoted $S(x,x)$.  As $S(x,x)$ requires a source at each lattice
point, a different recipe to that of conventional point-to-all
propagators is utilised.  For this purpose we use stochastic
estimation of the matrix inverse \cite{Dong:1993pk,Foley:2005ac}.

Given a set of random noise vectors $\{\eta\}$ with elements drawn
from $\mathbb{Z}_{4}$ such that the average over noise vectors gives
\begin{equation} \langle \eta_{a\alpha}(x) \, \eta_{b\beta}^\dagger(y) \rangle = \delta_{xy}\,\delta_{ab}\,\delta_{\alpha\beta}, \end{equation}
with colour indices $a,b$, spin indices $\alpha, \beta$ and space-time indices $x,y$.  We define for each noise vector a corresponding solution vector
\begin{equation} \chi = M^{-1}\,\eta, \end{equation}
where in this case $M$ is the fermion matrix. Then the stochastic
estimate of a propagator matrix element is calculated as
\begin{equation}
S_{{ab};\alpha\beta}(x,y) \simeq  \langle \chi_{a\alpha}(x)\,\eta_{b\beta}^\dagger(y) \rangle. \label{eq:StochEst}
\end{equation}
We perform full dilution in time, spin and colour indices as a means
of variance reduction \cite{O'Cais:2004ww}.  That is, given a set of
full noise vectors $\{\eta\},$ we can define a set of diluted noise vectors
$\{\eta^{[a'\alpha't']}\}$ by
\begin{equation}
\eta^{[a'\alpha't']}_{a\alpha}(\vec{x},t) = \delta_{aa'}\,\delta_{\alpha\alpha'}\,\delta_{tt^{\prime}}\,\eta_{a\alpha}(\vec{x},t), 
\end{equation}
where the intrinsic quark field indices are specified by colour $a,$
spin $\alpha,$ space $\vec{x}$ and time $t$ respectively and the
colour-spin-time diluted noise vectors are enumerated by the
corresponding $[a'\,\alpha'\,t']$ labels. We can similarly enumerate the
solution vectors
\begin{equation} \chi^{[a'\alpha't']} = M^{-1} \eta^{[a'\alpha't']}, \end{equation}
which makes it clear that by diluting we increase the number of
inversions required by a factor of $n_{\rm colour}\times n_{\rm
  spin}\times n_{\rm time}.$ The stochastic estimate of the matrix
inverse with dilution is given by
\begin{equation}
S(x,y) \simeq  \langle \sum_{a',\alpha',t'}\chi^{[a'\alpha't']}(x)\,\eta^{[a'\alpha't']\dagger}(y)\, \rangle,
\end{equation}
where colour and spin indices are taken to be implicit for clarity.
At this point we remark that, while it is computationally infeasible
to also fully dilute in the space index $\vec{x},$ in this extreme
limit each diluted noise vector would consist of only a single
non-zero element, meaning that we are exactly calculating the full
matrix $S(x,y)$ and the above relation becomes an equality rather than
an estimate. This makes it clear that using dilution provides an
improved stochastic estimate to the matrix inverse.

As shown in Fig.~\ref{fig:FeynmanDiag}, our construction of nucleon
correlators with five-quark operators combines standard point-to-all
propagators $S(x,0)$ and stochastic estimates of the loop propagators
$S(x,x).$ In order to access the radial excitations of the nucleon, we
make use of multiple levels of Gaussian smearing in our quark
fields. Hence, to construct a correlation matrix we need to calculate
propagators with differing levels of source and sink smearing. 

Let $S^{(m,n)}(x,y)$ denote a propagator with $m$ iterations of smearing
applied at the sink and $n$ iterations applied at the source. In the
case of point-to-all propagators $S^{(m,n)}(x,0)$ the source point is
fixed, $y=0,$ and starting with a point source $\psi^{(0)},$ we apply
$n$ iterations of Gaussian smearing pre-inversion to obtain the
smeared source $\psi^{(n)} = H^n \, \psi^{(0)},$ where
\begin{multline}
H\,\psi(x) = (1-\alpha)\,\psi(x) + \frac{\alpha}{6}\sum_{\mu=1}^3\left\{ U_\mu(x)\,\psi(x+a\hat{\mu})\right. \\
\left. + U^\dagger_\mu(x-a\hat{\mu})\,\psi(x-a\hat{\mu})\right\},\label{eq:Smear}
\end{multline}
and $\alpha$ specifies the smearing fraction. Sink smearing is applied
to the propagator post-inversion to obtain $S^{(m,n)}(x,0).$

The application of smearing to construct a stochastic estimate for the quark
propagator $S^{(m,n)}(x,y)$ is somewhat different. The set of (diluted) noise
and solution vectors $\{\eta,\chi\}$ is first constructed, whereby it
follows from Eqs.~(\ref{eq:StochEst}) and (\ref{eq:Smear}) that an
estimate of the smeared propagator is given by
\begin{equation} S^{(m,n)}(x,y) =  \langle \chi^{(m)}(x)\eta^{(n)\dagger}(y) \rangle, \end{equation}
where $\chi^{(m)} = H^m \chi$ is the result of $m$ iterations of
Gaussian smearing applied to the (diluted) solution vectors, and $\eta^{(n)}=H^n
\eta$ is similarly constructed from the (diluted) noise vectors. Note that the
smearing is applied \emph{after} (any dilution and) the solution vectors have been
calculated. The construction of a smeared loop propagator
$S^{(m,m)}(x,x)$ is simply an application of the above formulae in the
case $y=x.$

\begin{figure}[t]
  {\includegraphics[width=0.44\textwidth]{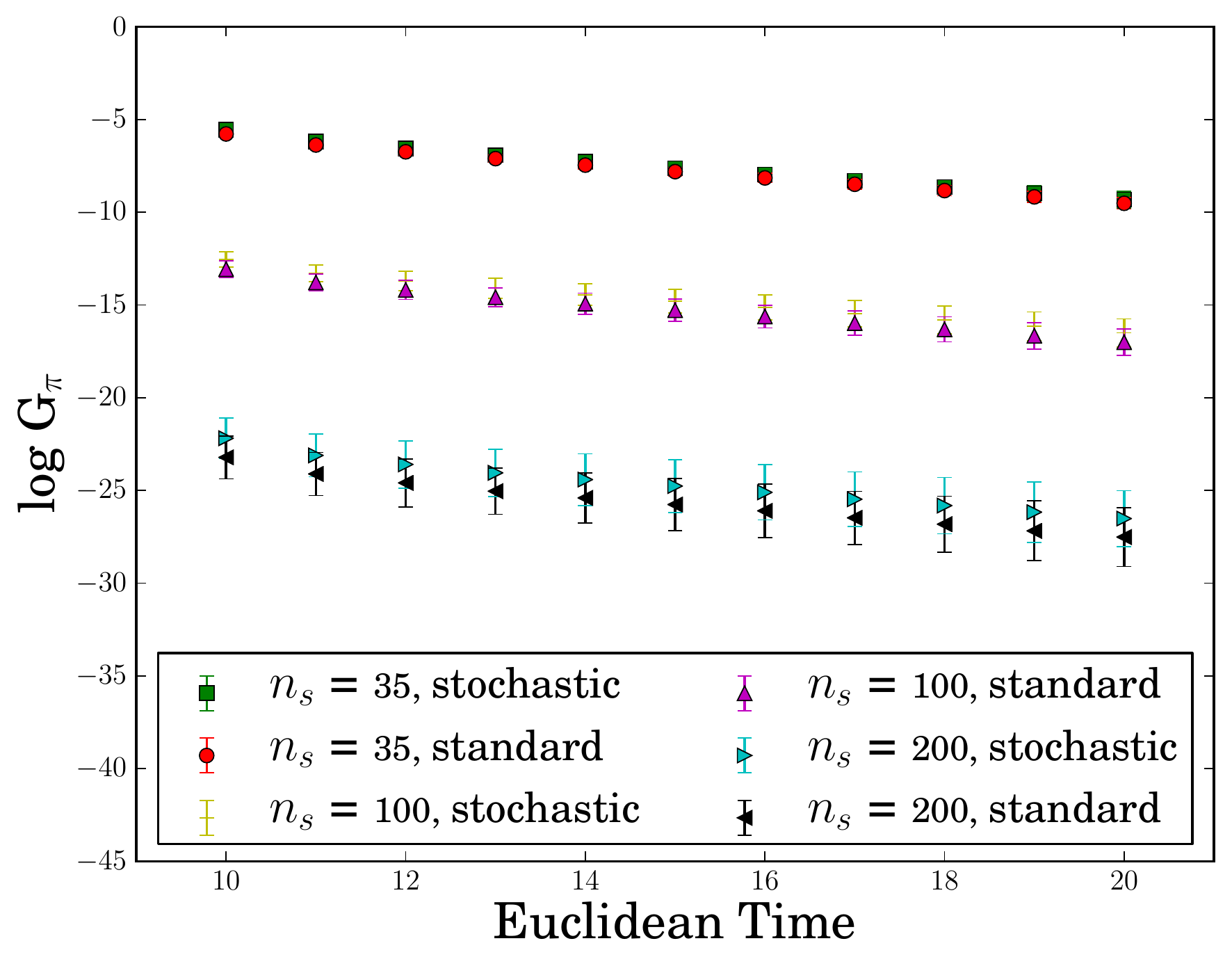}}\\
  {\includegraphics[width=0.44\textwidth]{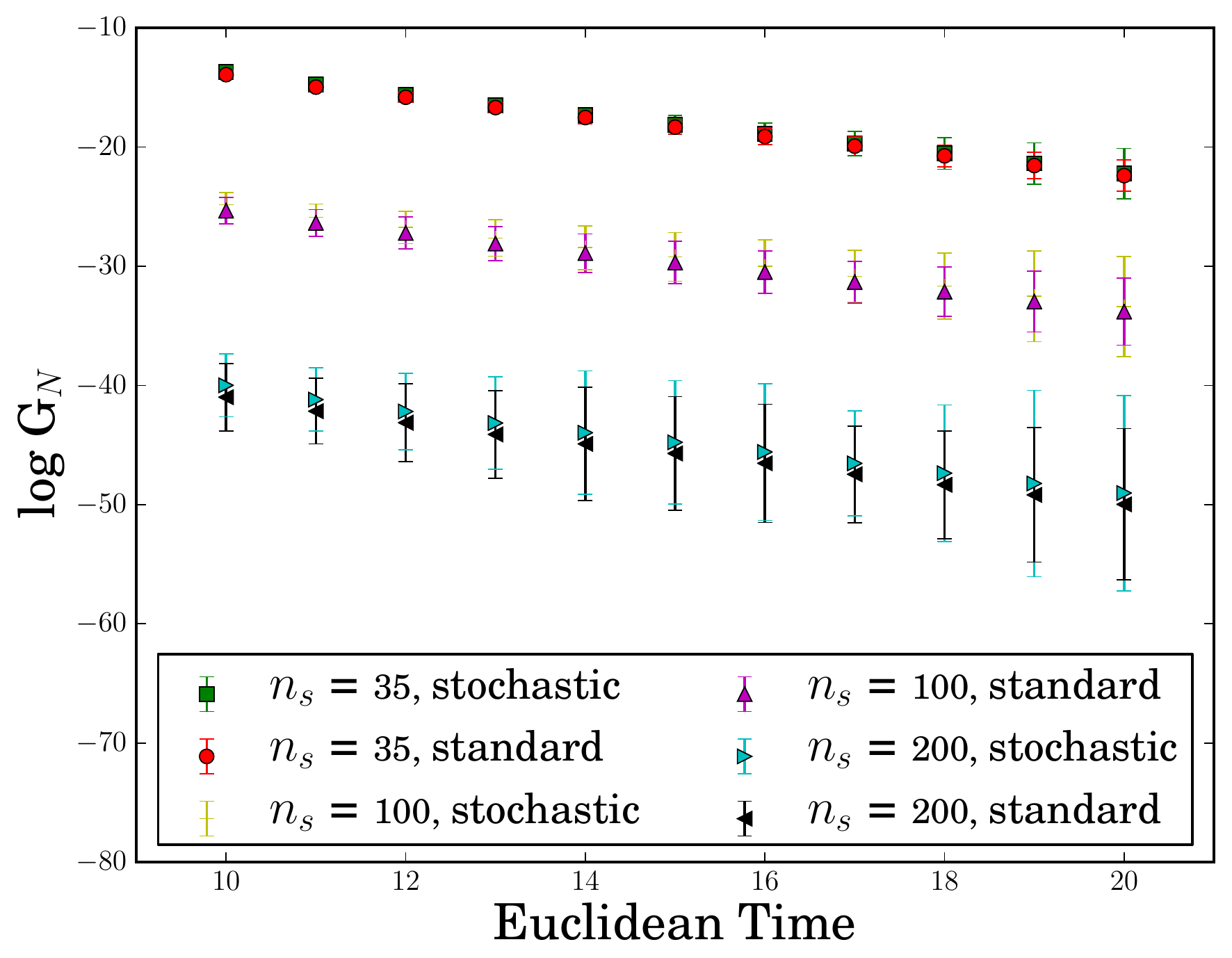}}
 \caption{(Colour online).  A comparison of correlators calculated with one stochastically estimated propagator (denoted ``stochastic'') to those calculated with no stochastic propagators (denoted ``standard'').  Results are presented for the pion (top) and the ground state nucleon (bottom).}
  \label{fig:StochCompare}
\end{figure}

In order to determine how many noise vectors per configuration are
sufficient to provide similar statistical errors for our point-to-all
and stochastic propagators, correlators are calculated with a
stochastic estimate of the point-to-all propagator, and compared to
those obtained using point-to-all propagators calculated in the
standard way. As each independent quark line in a hadron correlator
requires an independent noise source to ensure unbiased
estimation~\cite{Morningstar:2011ka} we insert one stochastic
propagator into the aforementioned correlators.  Furthermore,
as a test of our smearing technique for stochastic propagators, we
perform this comparison using a variety of smearing levels. Note that,
as smearing of both the source and solution vectors is performed
post-inversion, the stochastic method effectively provides different
smearing levels for free.  

The comparison is performed on 75 $20^3 \times 40$ lattice gauge configurations, with the
FLIC fermion action~\cite{Zanotti:2001yb}.  The lattice spacing is
$0.126\textrm{ fm}$ in both the temporal and spatial direction
providing a physical lattice volume of $(2.52\textrm{ fm})^3$.  Four
full noise vectors are used per stochastic propagator, which are then
colour-spin-time diluted.  As the source timeslice is fixed in this
case, each stochastic propagator requires $n_{\rm colour}
\times n_{\rm spin}$ inversions per noise vector.  Recall standard point-to-all propagators require $n_{\rm colour}
\times n_{\rm spin}$ inversions, although source smearing is applied pre-inversion unlike the stochastic case.  Three different levels of smearing
are used, $n_s=35,100,200$ sweeps with $\alpha=0.7.$ Fig.
\ref{fig:StochCompare} shows good agreement across all smearing levels
between those correlators containing a stochastically estimated
propagator and those that do not, demonstrating that using four noise
vectors per quark line provides a comparable statistical uncertainty
to that of a standard propagator. We note here that ultimately we
utilise this method to calculate $S(x,x)$ not $S(x,x_{source}),$
meaning we get the added benefit of spatial averaging for our loop
propagators.
%
%
\section{Simulation Details}
\label{sect:Simulation Details}
For the baryon spectroscopy results presented herein we use the PACS-CS $2 + 1$ flavour
dynamical-fermion configurations~\cite{Aoki:2008sm} made available
through the ILDG~\cite{Beckett:2009cb}.  These configurations use the
non-perturbatively $\mathcal{O}(a)$-improved Wilson fermion action and
the Iwasaki gauge action~\cite{Iwasaki:1983ck}.  The lattice size is
$32^3 \times 64$ with a lattice spacing of $0.0907\textrm{ fm}$
providing a physical volume of $\approx (2.90\textrm{ fm})^3$.  $\beta
= 1.90$, the light quark mass is set by the hopping parameter
$\kappa_{ud} = 0.13770$ which gives a pion mass of $m_{\pi} = 293
\textrm{ MeV}$, while the strange quark mass is set by $\kappa_{s} =
0.13640$.
Fixed boundary conditions are employed in the time direction removing backward propagating states~\cite{Melnitchouk:2002eg, Mahbub:2009nr}, and
the source is inserted at $t_{src} = n_t/4 = 16$, well away from the boundary.  Systematic effects associated with this boundary condition are negligible for $t > 16$ slices from the boundary. 
The main results of our
variational analysis is performed at $t_{0}=17$ and $\dt=3,$
providing a good balance between systematic and statistical
uncertainties.  Uncertainties are obtained via single elimination
jackknife while a full covariance matrix analysis provides the
$\chi^{2}/dof$ which is utilised to select fit regions for the
eigenstate-projected correlators.

In addition to the five quark operators $\chi_{5}$ and $\chi_{5}^{\prime}$ presented in Section \ref{sect:MultiParticleStateContributions} we use the conventional three-quark operators
\begin{align}\label{NucleonInterps}
\chi_{1} &= \epsilon^{abc}[u^{aT}\, (C\gamma_{5})\, d^{b}]\, u^{c}\nonumber\\
\chi_{2} &= \epsilon^{abc}[u^{aT}\, (C)\, d^{b}]\, \gamma_{5}\, u^{c}
\end{align}
in order to form the seven bases we study that are outlined in Table \ref{table:BasisTable}.
\begin{table}[!h]
\caption{Table of the various operators used in each basis.}
\begin{ruledtabular}
\begin{tabular}{cc}
    Basis Number & Operators Used  \\[2pt]
    \hline 
\noalign{\vspace{2pt}}
    1 & $\chi_{1}$, $\chi_{2}$\\
    2 & $\chi_{1}$, $\chi_{2}$, $\chi_{5}$\\
    3 & $\chi_{1}$, $\chi_{2}$, $\chi_{5}^{\prime}$\\
    4 & $\chi_{1}$, $\chi_{2}$, $\chi_{5}$, $\chi_{5}^{\prime}$\\
    5 & $\chi_{1}$, $\chi_{5}$, $\chi_{5}^{\prime}$\\
    6 & $\chi_{2}$, $\chi_{5}$, $\chi_{5}^{\prime}$\\
    7 & $\chi_{5}$, $\chi_{5}^{\prime}$\\[2pt]
\end{tabular}
\end{ruledtabular}

\label{table:BasisTable}
\end{table}

Throughout this work we employ Gauge-invariant Gaussian
smearing~\cite{Gusken:1989qx} at the source and sink to increase the
basis size via altering the overlap of our operators with the states
of interest.  We choose $n_s = 35$ and $n_s = 200$ sweeps of smearing
providing bases of sizes 4, 6 and 8. Stochastic quark lines are
calculated using four random $\mathbb{Z}_4$ noise vectors that are
fully diluted in colour, spin and time.

\section{Results}
\label{sect:Results}

\subsection{Positive-parity Spectrum}

\begin{figure}[htbp]
  \centering 
  {\includegraphics[width=0.48\textwidth]{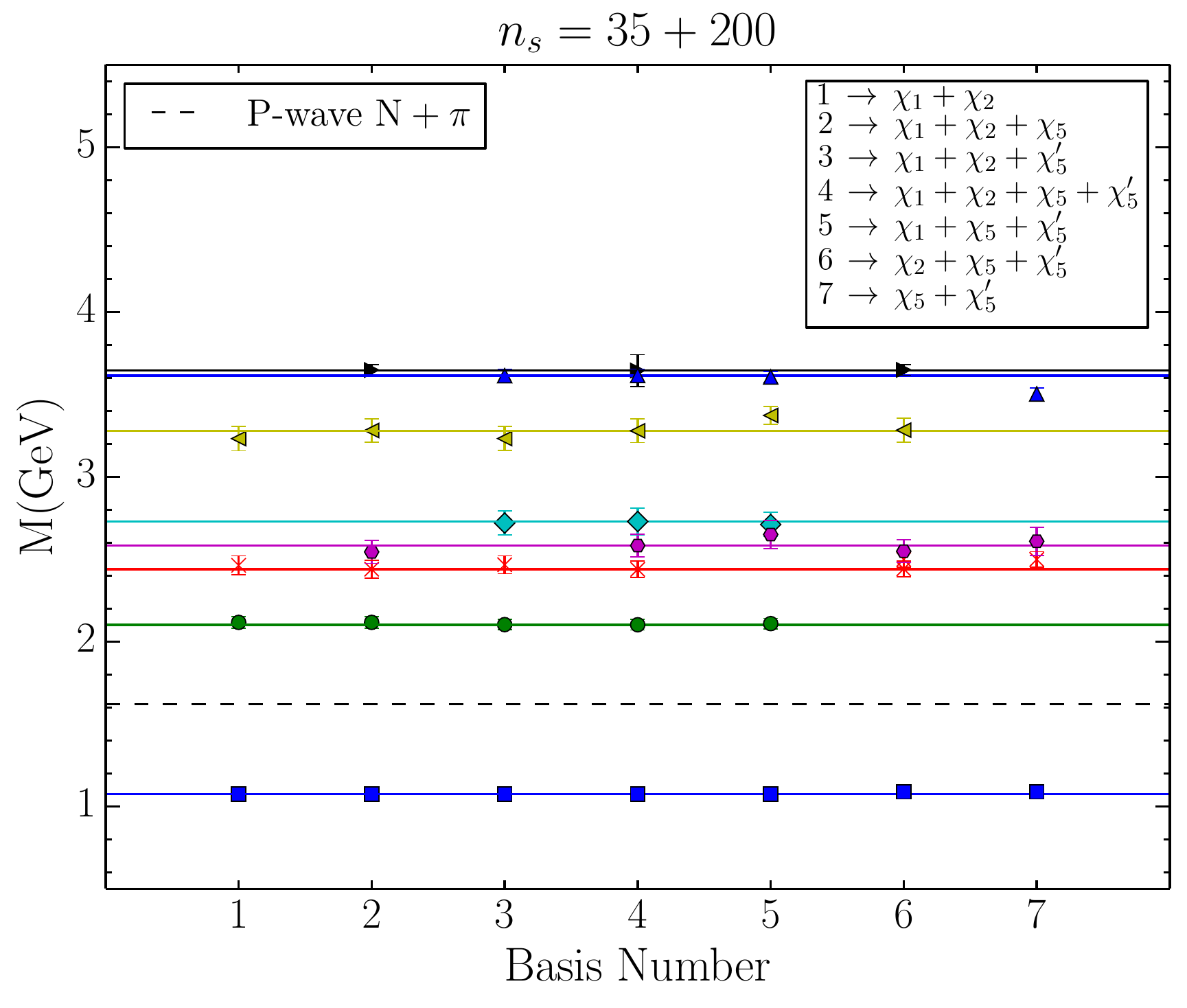}}
 \caption{(Colour online).  The positive-parity nucleon spectrum with various operator bases constructed with 35 and 200 sweeps of smearing.  Horizontal solid lines are present to guide the eye and are drawn from the central value of the states in basis 4, while the dashed line marks the position of the non-interacting P-wave $N\pi$ scattering threshold}.
 \label{fig:Masses+}
\end{figure}

The results for the nucleon spectrum in the positive-parity sector are
shown in Fig.~\ref{fig:Masses+}. Solid horizontal lines are added to
guide the eye, with their values set by the states in basis number 4,
since this basis contains all the operators studied and has the
largest span.

Of particular interest is the robustness of the variational techniques
employed.  While changing bases may effect whether or not a particular
state is seen, the energy of the extracted states is consistent across
the different bases, even though they contain qualitatively different
operators.

\begin{figure}[htbp]
  \centering 
  {\includegraphics[width=0.48\textwidth]{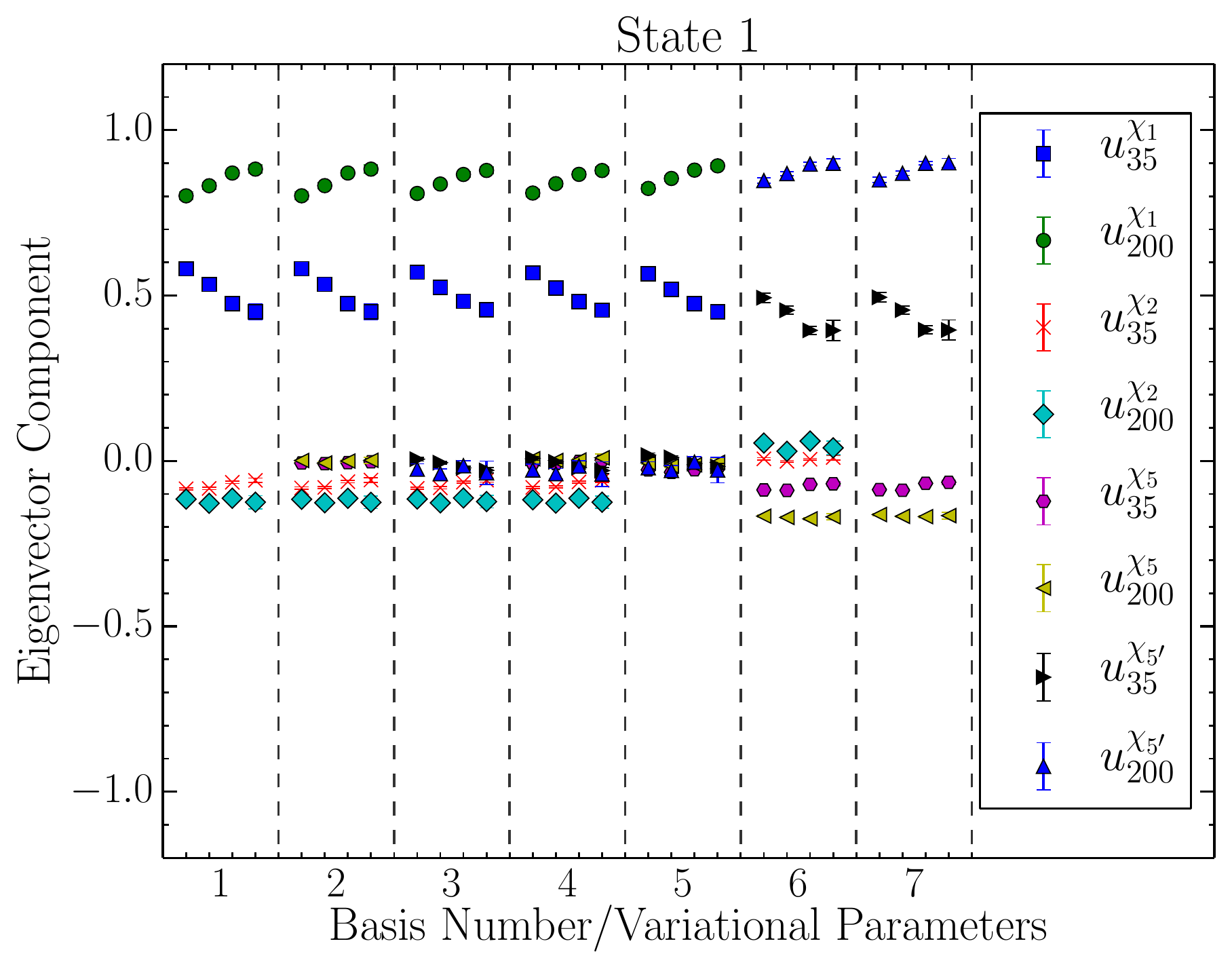}}\\
  {\includegraphics[width=0.48\textwidth]{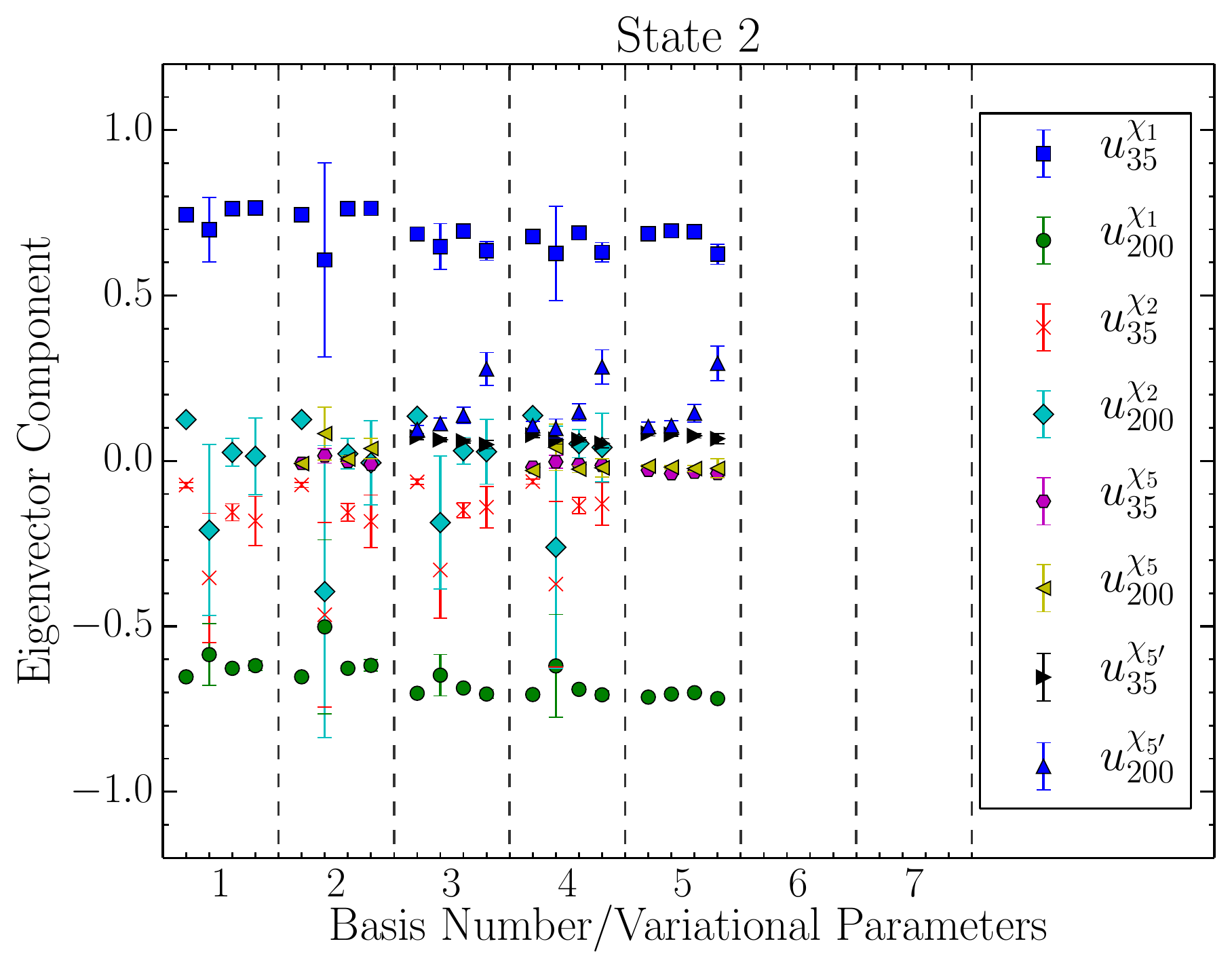}}\\
  {\includegraphics[width=0.48\textwidth]{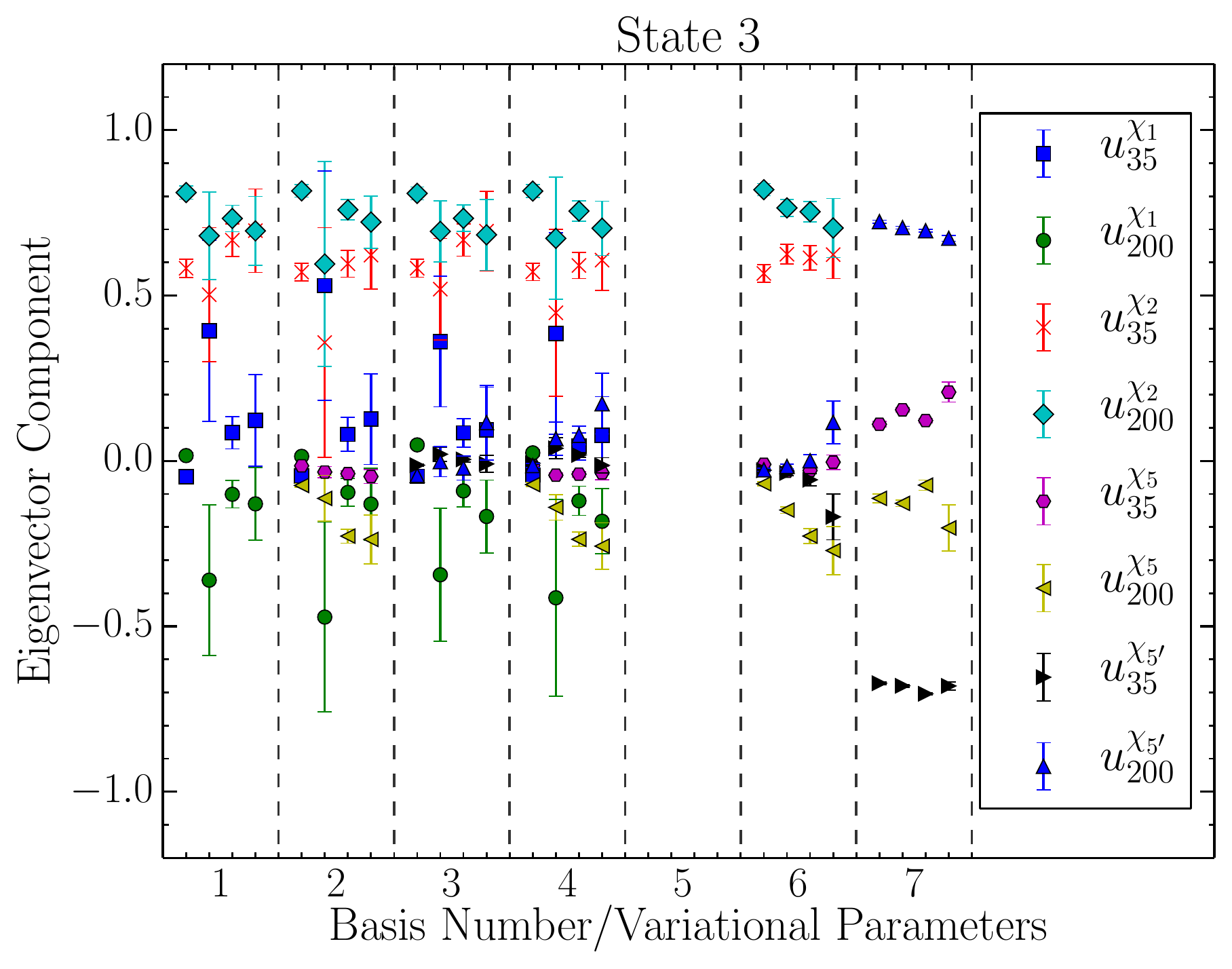}}
  \caption{(Colour online).  Eigenvector components corresponding to
    the low-lying positive-parity nucleon states. State 1 corresponds
    to the ground state, with states 2 and 3 corresponding to the
    first and second excited states respectively. The column numbers
    denote basis number while the minor $x$ axis ticks correspond to
    the values of the variational parameter $\dt$ which runs from
    1 through to 4.  $t_{0} = 17$ has been used throughout.  The subscripts 35 and 200 in the legend refer to the number of smearing sweeps applied.}
 \label{fig:Evectors+}
\end{figure}

Despite the use of 5-quark operators, no state near the non-interacting
P-wave $N\pi$ scattering threshold is observed.
This is understood by noting that none of our operators have a source
of the back-to-back relative momentum between the nucleon and pion necessary to observe an energy level in the region of this scattering state.

The corresponding eigenvector components for the positive-parity
states are shown in Fig.~\ref{fig:Evectors+} as a function of basis
and variational parameter $\dt,$ with $t_0=17$ fixed. The values of
$dt$ range from 1 through 4. The upper limit of $\dt=4$ was chosen as
the largest value for which the variational analysis converged for each
of the seven bases.

The ground-state nucleon is observed in every basis regardless of the
absence or presence of a particular operator. If $\chi_1$ is present
then this provides the dominant contribution, with $\chi_{5}^{\prime}$
coupling strongly to the ground state in bases where $\chi_1$ is
absent.  An interesting interplay between 35 and 200 sweep smeared $\chi_1$ is observed with the smaller source diminishing in importance as $dt$ is increased.  This may be associated with the Euclidean time evolution of highly excited states which are suppressed with increasing $dt$.

Turning our attention to state 2, we see that $\chi_{1}$ plays a critical role in the extraction of the first excited state, which is associated with a radial excitation of
the ground state~\cite{Roberts:2013oea}.  Here the 35 and 200 sweep $\chi_1$ interpolators enter with similar strength but opposite signs, setting up the node structure of a radial excitation.  $\chi_{1}$ dominates the
construction of the optimised operator for this state for bases 1
through 5, whereas basis 6 and 7 which lack $\chi_{1}$ do not observe
this state.

The eigenvectors for state 3, the second excited state, are dominated
by $\chi_2$ components with the same sign when this operator is
present (bases 1-4,6). This state is not observed in basis 5 (where
$\chi_2$ is absent). Interestingly, in basis 7 which only contains
five-quark operators it appears that it is possible to form this state
using $\chi_5'$ components at two different smearings with opposite
sign.

We observe that the overall structure of the eigenvectors for each of
the three states is highly consistent across different bases and
different values of the variational parameter $\dt.$ The
structure of the eigenvectors can be considered to be a signature or
fingerprint of the extracted state, and this consistency across bases
confirms that it is the same state being identified. 

It is fascinating to see that for state 1 in bases 6 and 7, where $\chi_5'$
takes the role of the absent $\chi_1$ operator, the values of the two
dominant eigenvector components (which indicate the mixture of the two
different smearing levels used) are extremely similar to the $\chi_1$
components in bases 1-5. Interestingly, at $\dt=2$ the error bars for
the dominant components of states 2 and 3 blow up. As we shall explain
below, this is due to an accidental degeneracy in the eigenmasses for
this choice of variational parameters.

\begin{figure}[htbp]
  \centering 
  {\includegraphics[width=0.48\textwidth]{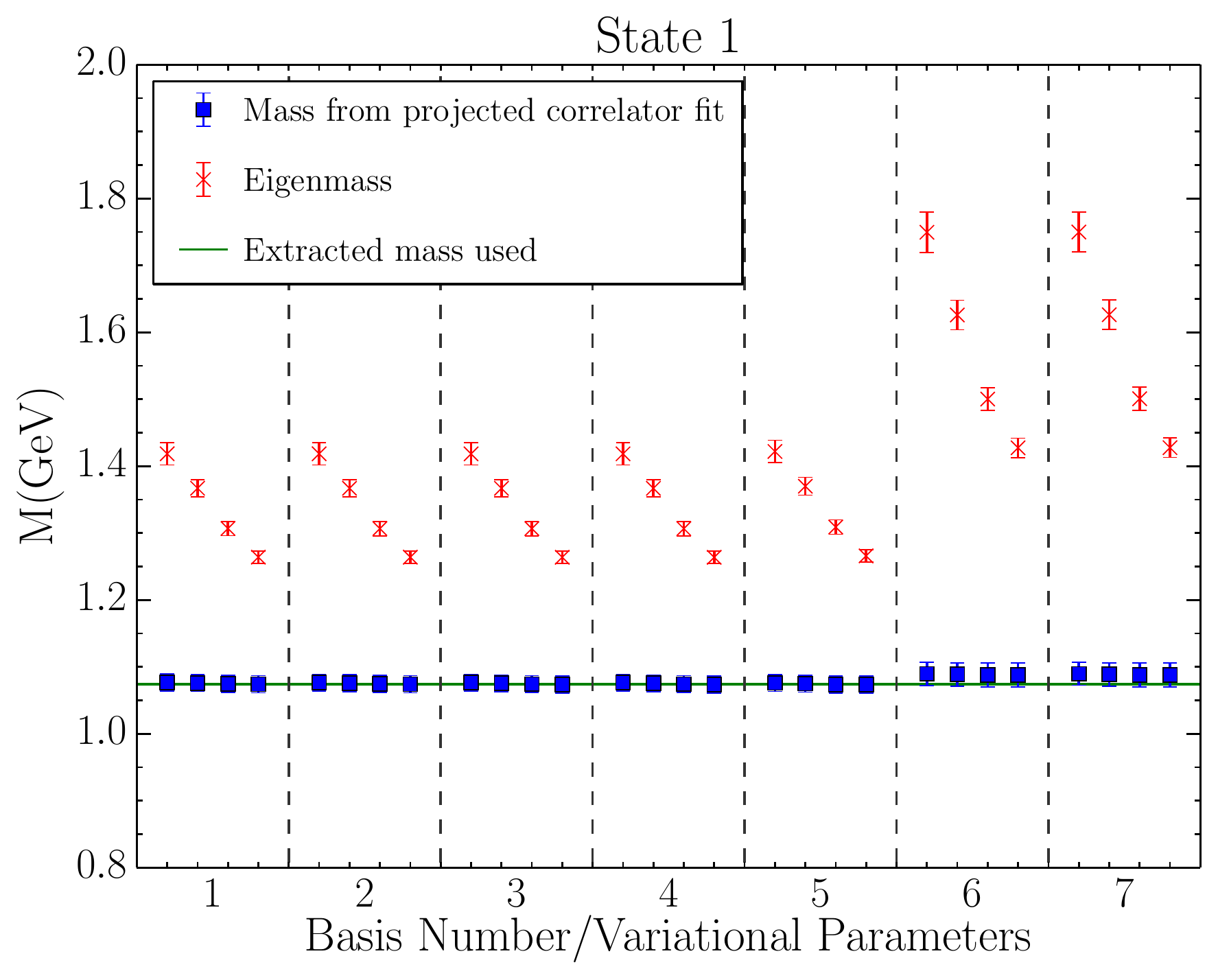}}\\
  {\includegraphics[width=0.48\textwidth]{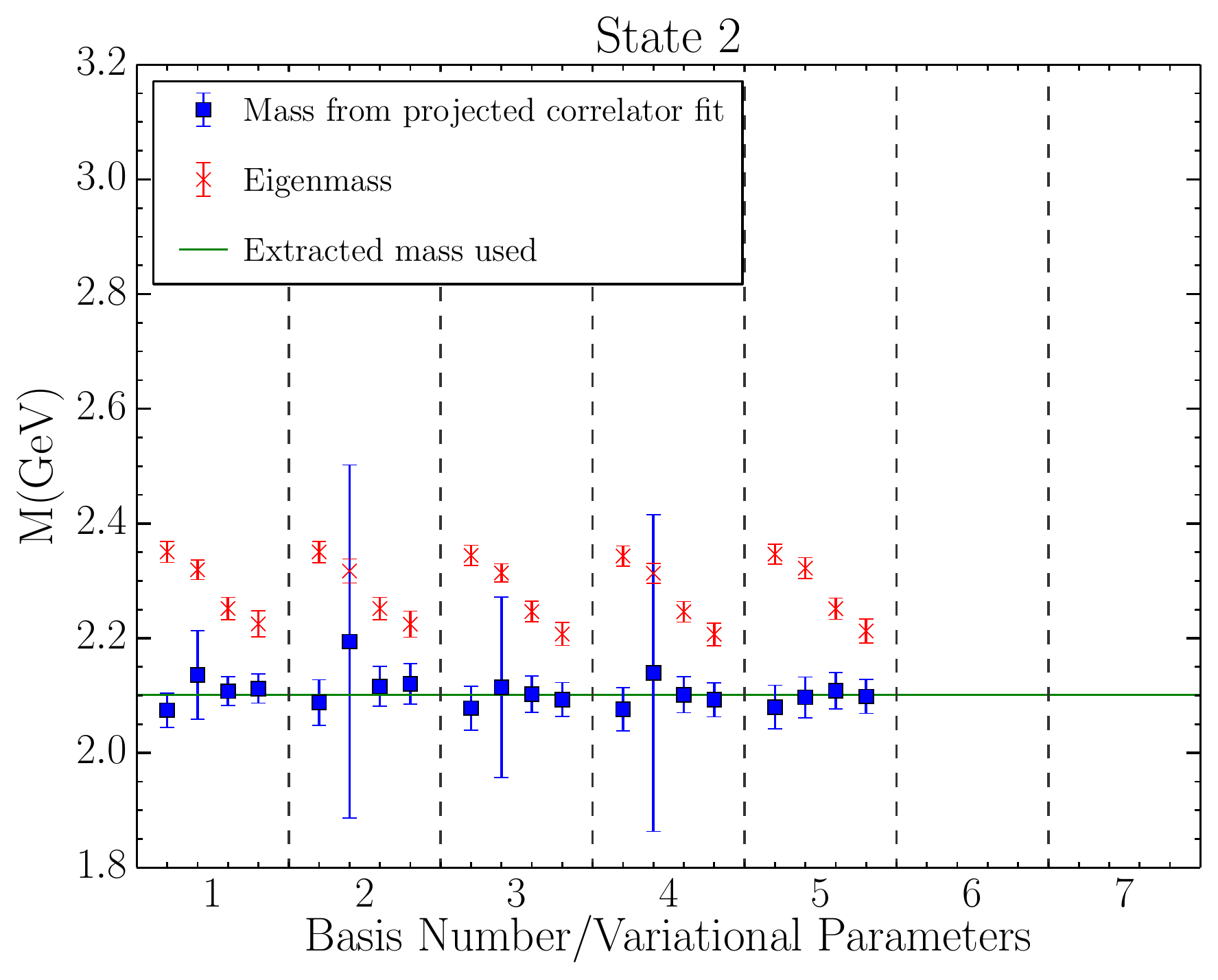}}\\
  {\includegraphics[width=0.48\textwidth]{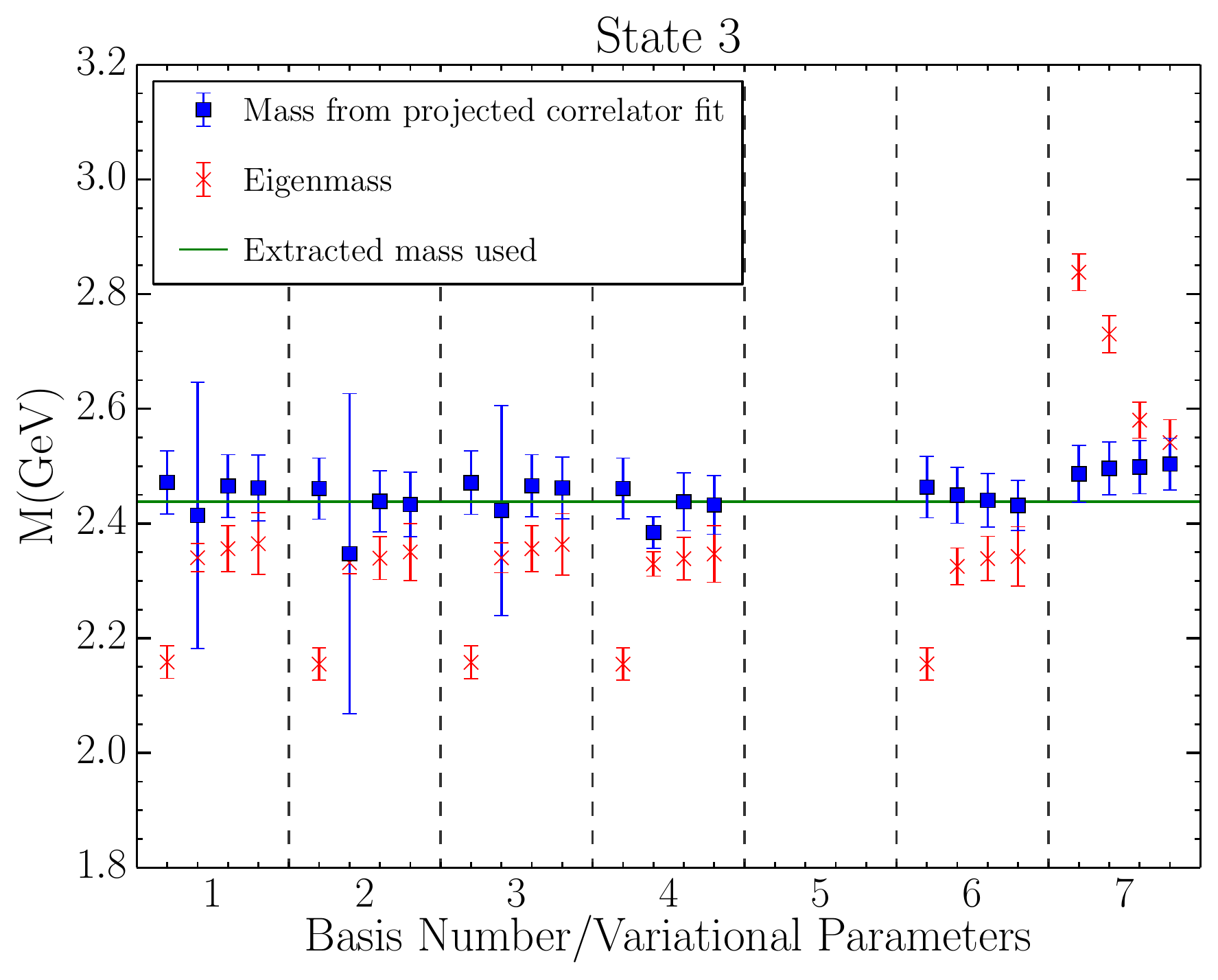}}
  \caption{(Colour online).  Comparisons of eigenmasses to masses obtained from a projected correlator fit for low-lying states in the positive-parity nucleon channel. The column numbers denote basis number while the minor $x$ axis ticks correspond to the values of the variational parameter $\dt=1\ldots 4.$  $t_{0} = 17$ has been used throughout.  The line denoting the extracted mass is set using basis 4 with $\dt = 3.$}
 \label{fig:EmassvsProjMass+}
\end{figure}

In order to further test the robustness of our variational method we
conduct a comparison of the masses obtained from fitting the
eigenstate-projected correlators as a function of the variational
parameters for each basis.  These results are presented in
Fig.~\ref{fig:EmassvsProjMass+}. Also shown for comparison are the
eigenmasses, $m_{\alpha}$, that result from solving the generalised eigenvalue equation of Eqs. (\ref{E-value-eq-L}) or (\ref{E-value-eq-R}) with $c^{\alpha} = \textrm{e}^{-m_{\alpha}\dt}$.

Studying state 1, the nucleon ground state, we observe that the masses
obtained from projected correlator fits are approximately invariant
across different bases and choices of the variational parameter. In
contrast, the eigenmass lies well above the fitted mass, dropping in
value as the variational parameter $\dt$ is varied from 1 to
4. While the eigenmass is directly related to the principal
correlator and thus should approach the ground state mass in the
large time limit, it is clear that the values of $\dt$ we examine here
are insufficient for this to occur.  It is worth noting that, in bases
6 and 7 where $\chi_1$ is absent we see that the eigenmass value rises
significantly.  Nevertheless, the fitted mass remains
remarkably consistent with the values obtained in bases 1-5.  We emphasize how strong the variational parameter dependence of the eigenmass contrasts the more consistent structure of the eigenvectors.  Insensitivity of the eigenvectors to the variational parameters is a key component of the invariance of the masses obtained from the projected correlator.

\begin{figure}[htbp]
  \centering 
  {\includegraphics[width=0.48\textwidth]{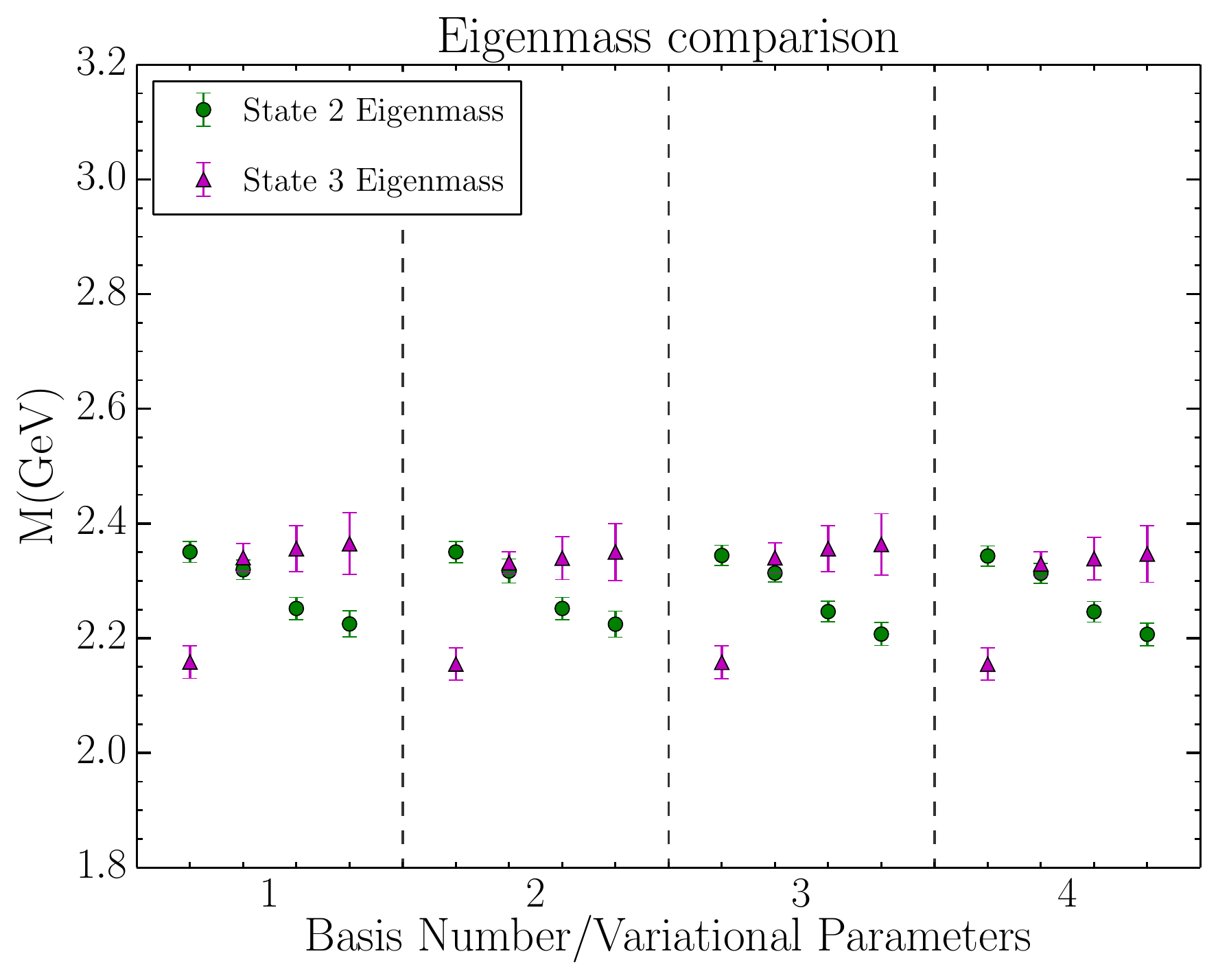}}
  \caption{(Colour online).  A plot showing the eigenmasses for both states 2 and 3, illustrating the accidental degeneracy at $dt=2$.}
 \label{fig:DegenerateEmass}
\end{figure}
Turning to state 2, we see that the eigenmass shows similar behaviour
to state 1, lying above the extracted mass and dropping with $\dt.$
Interestingly, for state 3 in bases 1-4 and 6 the eigenmass shows
constant behaviour for $\dt=2-4$ but systematically lies below the
extracted mass. In basis 7, the state 3 eigenmass is very different to
the previous bases, lying above the extracted mass and showing a
similar downward trend to states 1 and 2 as $\dt$ varies.

As for state 1, the fitted masses for states 2 and 3 provide highly
consistent values and uncertainties across the different bases and
values of $\dt,$ with the notable exception of $\dt=2.$ As observed
previously in Fig.~\ref{fig:Evectors+}, we see in
Fig.~\ref{fig:EmassvsProjMass+} considerably larger error bars at
the variational parameter set $(t_0,\dt) = (17,2)$ in both the
eigenvector components and projected mass fits for the first and
second excited states.  To understand this, we turn to Fig.~\ref{fig:DegenerateEmass}, where the eigenmasses for
states 2 and 3 are plotted against the variational parameter $\dt$ in
each basis.  

Note that at $(t_0,\dt) = (17,2)$ there is an approximate degeneracy
in the eigenmass for states 2 and 3. As a consequence, the
corresponding eigenvectors can therefore be arbitrarily rotated within
the state 2/state 3 subspace while remaining a solution to the
eigenvalue problem. When constructing the jackknife sub-ensembles to
calculate the error in the fitted energy, we need to solve for the
eigenvectors on each sub-ensemble. Due to the approximate degeneracy,
the particular linear combination of state 2 and state 3 that we
obtain for each sub-ensemble can vary. Indeed, we observe that the
dot-product between the ensemble average and sub-ensemble can drop
significantly for $\dt=2$ in comparison to other values of $\dt.$ This
causes a large variation in the sub-ensemble eigenvector components
and a correspondingly large error bar.  The simplest way to avoid the
problem of this accidental degeneracy is to select a different value
of the variational parameter.

\subsection{Negative-parity Spectrum}

\begin{figure}[htbp]
  \centering
  {\includegraphics[width=0.48\textwidth]{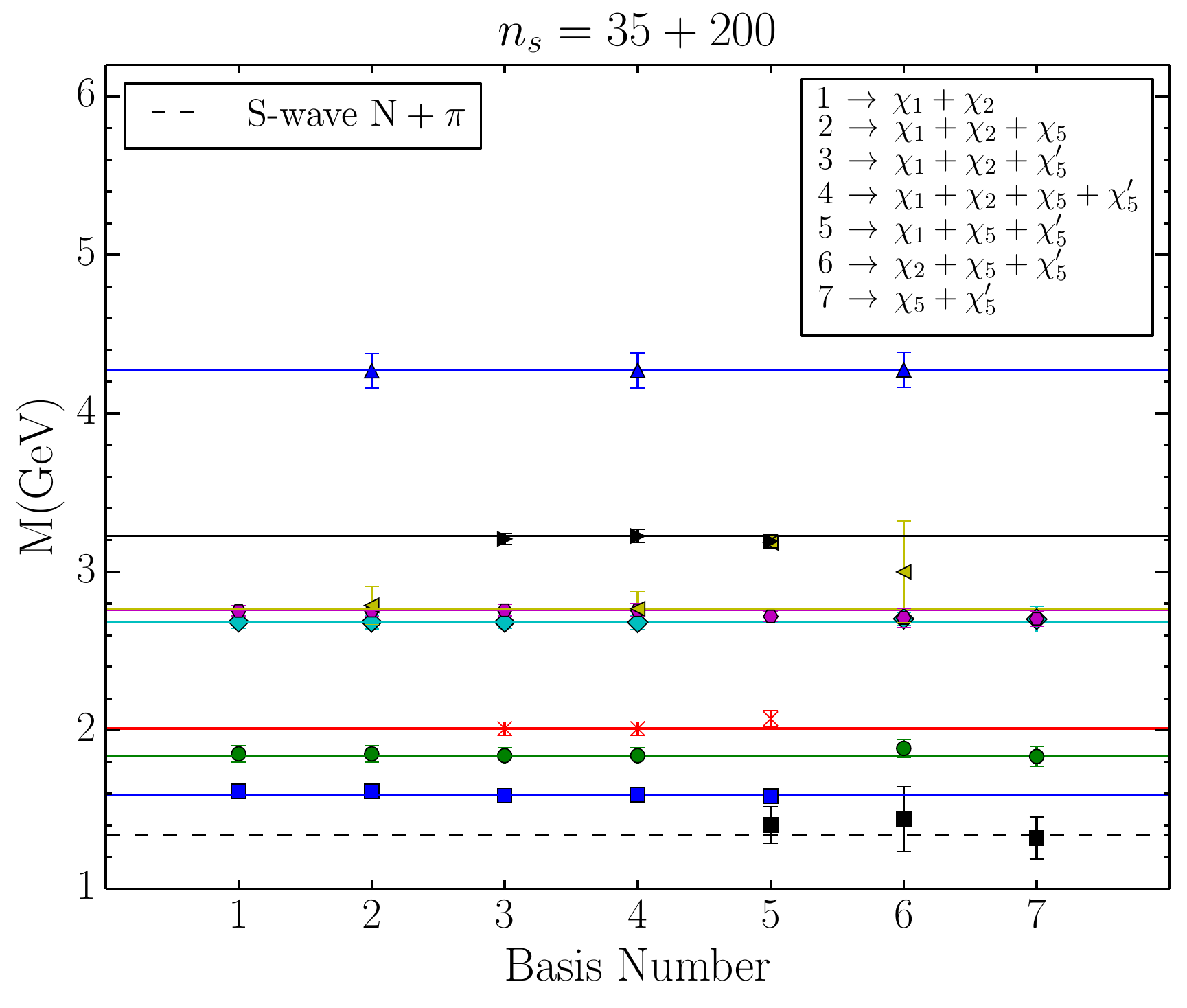}}
 \caption{(Colour online).  The negative-parity nucleon spectrum with various operator bases using 35 and 200 sweeps of smearing.  Solid horizontal lines are present to guide the eye and are drawn from the central value of the states in basis 4, since this basis is the largest.  The dashed line marks the position of the non-interacting S-wave $N\pi$ scattering threshold.  The variational parameters used herein are $(t_{0},\dt) = (17,3)$.}
 \label{fig:Masses-}
\end{figure}

The negative-parity nucleon spectrum is presented in Fig.
\ref{fig:Masses-}. Solid horizontal lines have been added to guide the
eye, with their values set by the states in the largest basis (number
4). Once again, while changing bases effects whether or not we observe a
given state, the extracted states display an impressive level of
consistency across the different bases.  

The dashed line indicating the energy of the non-interacting (infinite-volume) scattering-state threshold is also indicated with the caution that mixing with nearby states in the finite volume can alter the threshold position \cite{Hall:2013qba, Menadue:2011pd}.
We note here that all scattering thresholds discussed in this section and the next, refer to the non-interacting threshold.
In contrast to the positive-parity results, we do observe a state near the S-wave $N\pi$ scattering threshold in the negative-parity
channel (bases 5,6,7), also noting that the P-wave
$N\pi\pi$ thresholds lie in the region of state 3 seen in bases 3, 4 and 5.  It is important to note that even
after the introduction of operators that permit access to a state near the low-lying
scattering state, the energies of the higher states in the spectrum are consistent, demonstrating the robustness of the variational
techniques employed.

\begin{figure}[htbp]
  \centering
  {\includegraphics[width=0.48\textwidth]{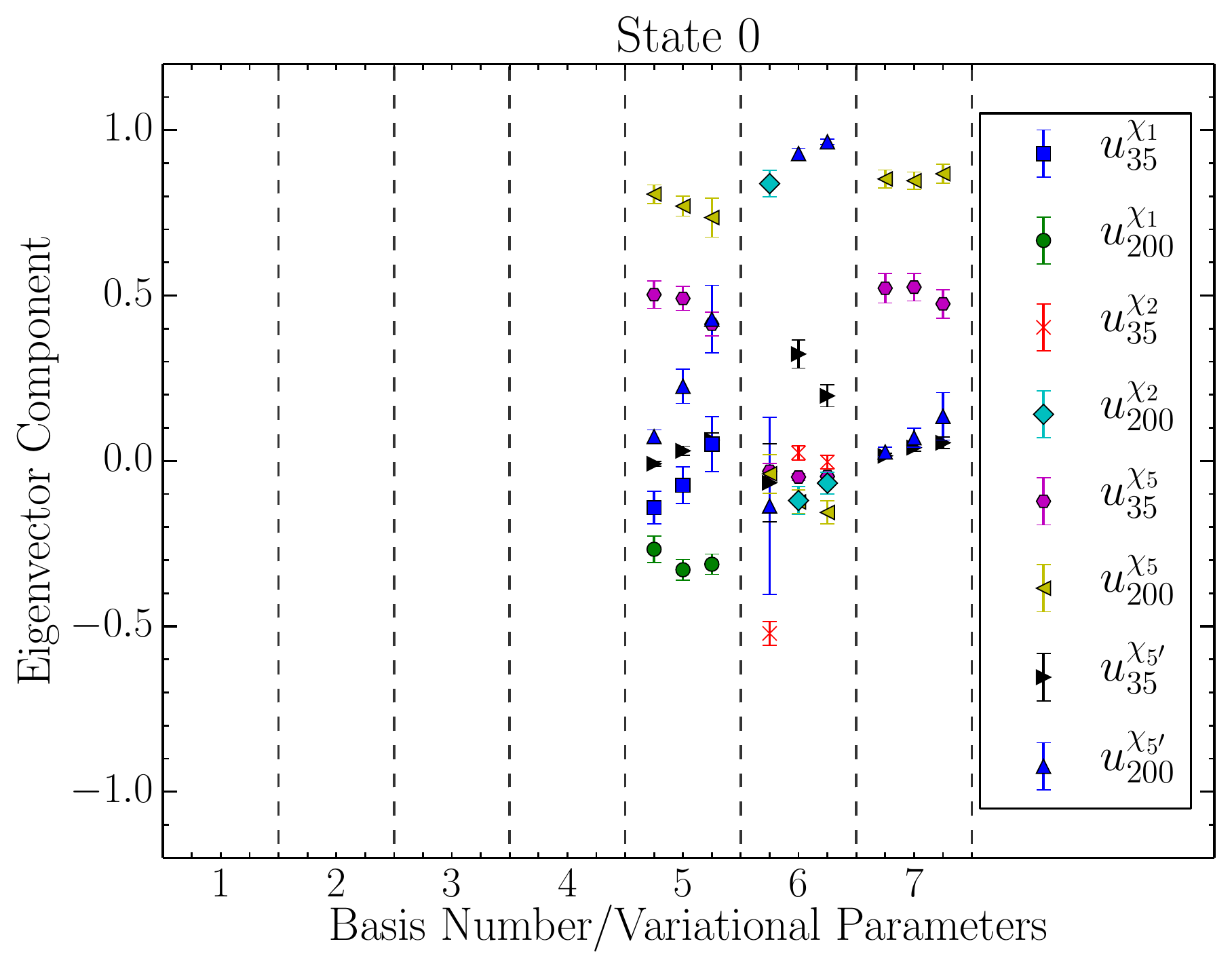}}\\
  \caption{(Colour online).  Eigenvector components corresponding to State 0 which is in the region of the non-interacting S-wave $N + \pi$ scattering threshold.  The column numbers denote basis number while the minor $x$ axis ticks correspond to the values of the variational parameter $\dt$ which runs from 1 through to 3.  $t_{0} = 17$ has been used throughout.  The subscripts 35 and 200 in the legend refer to the number of smearing sweeps applied.}
 \label{fig:Evectors-s0}
\end{figure}
\begin{figure}[htbp]
  \centering
  {\includegraphics[width=0.48\textwidth]{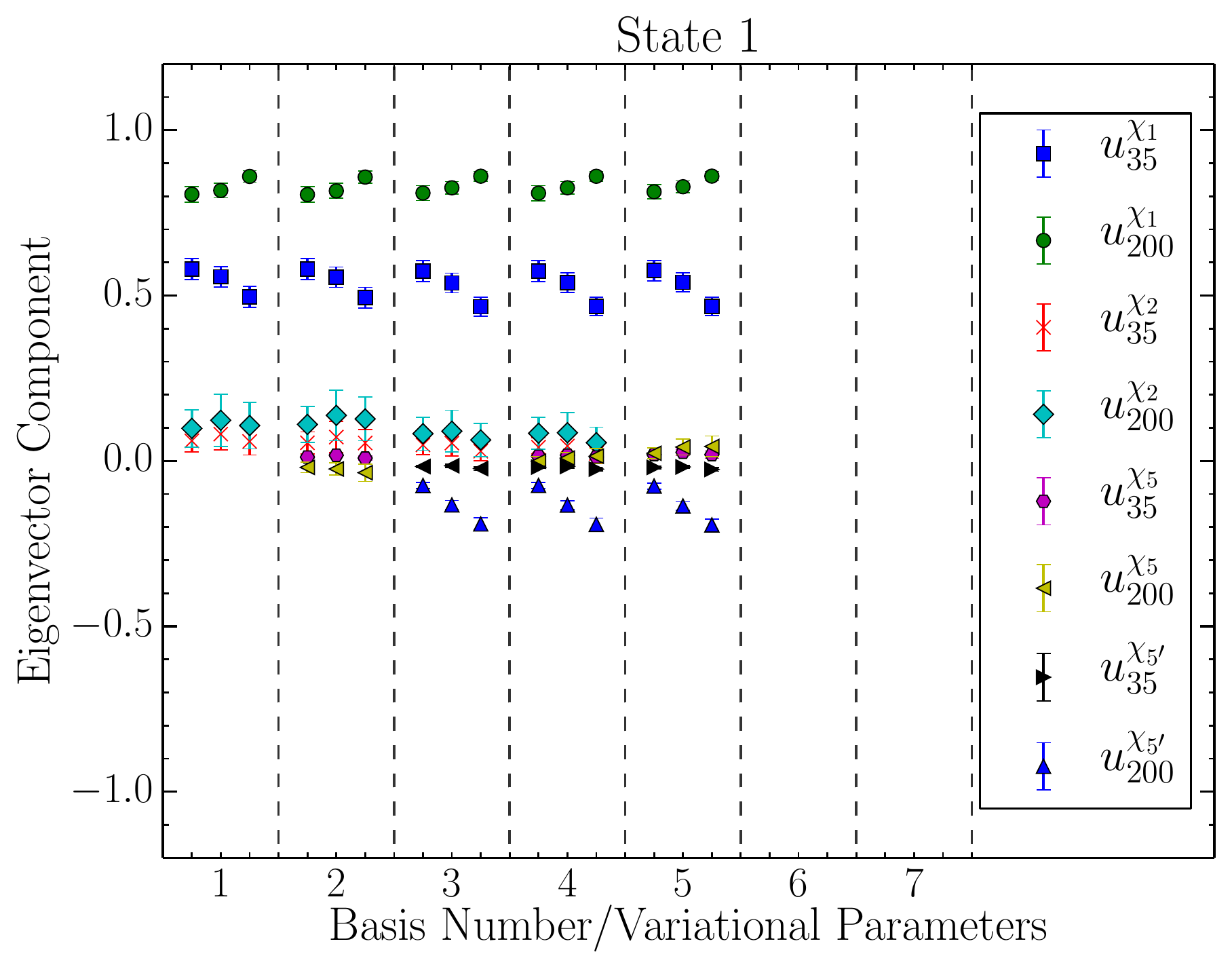}}\\
  {\includegraphics[width=0.48\textwidth]{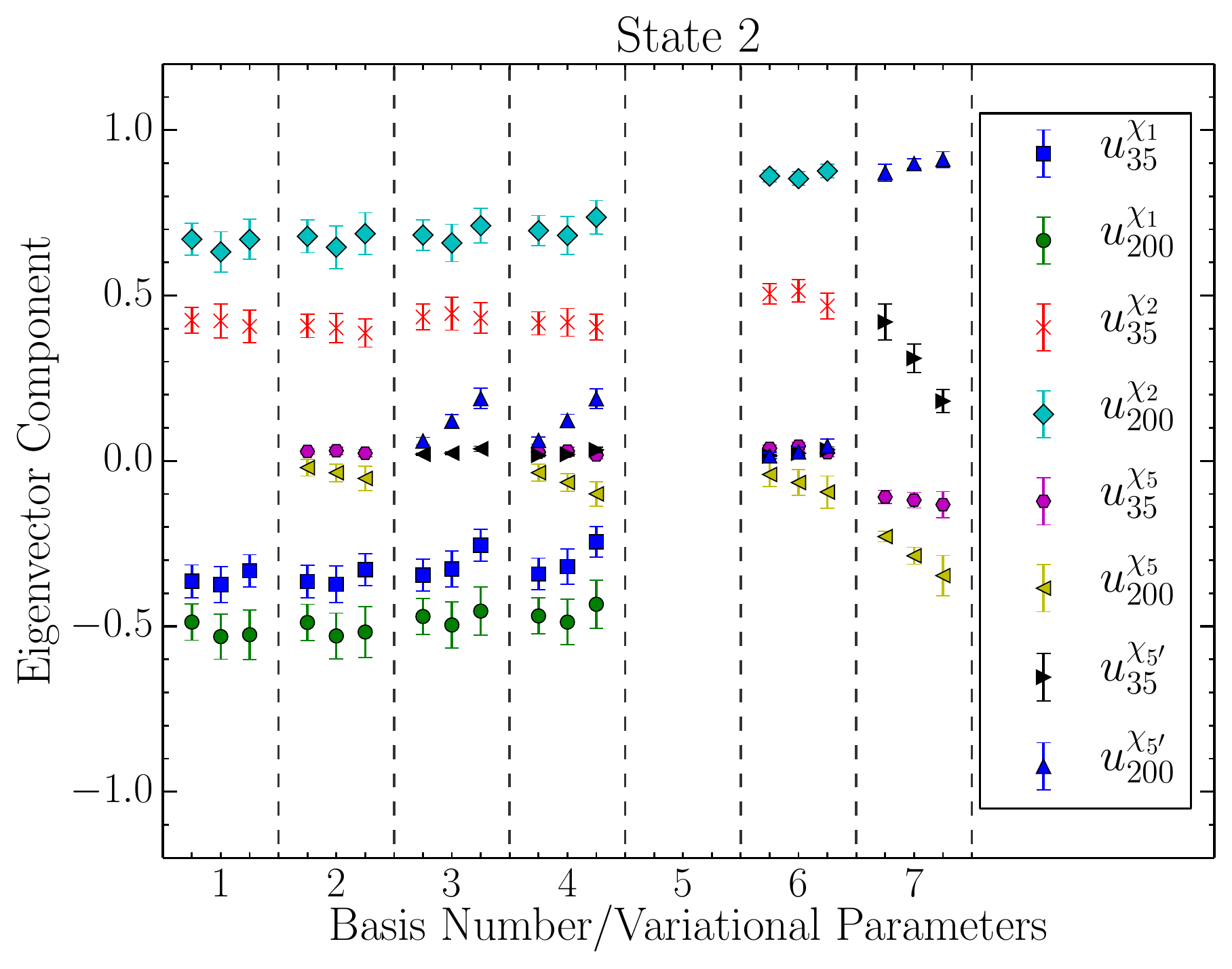}}\\
  {\includegraphics[width=0.48\textwidth]{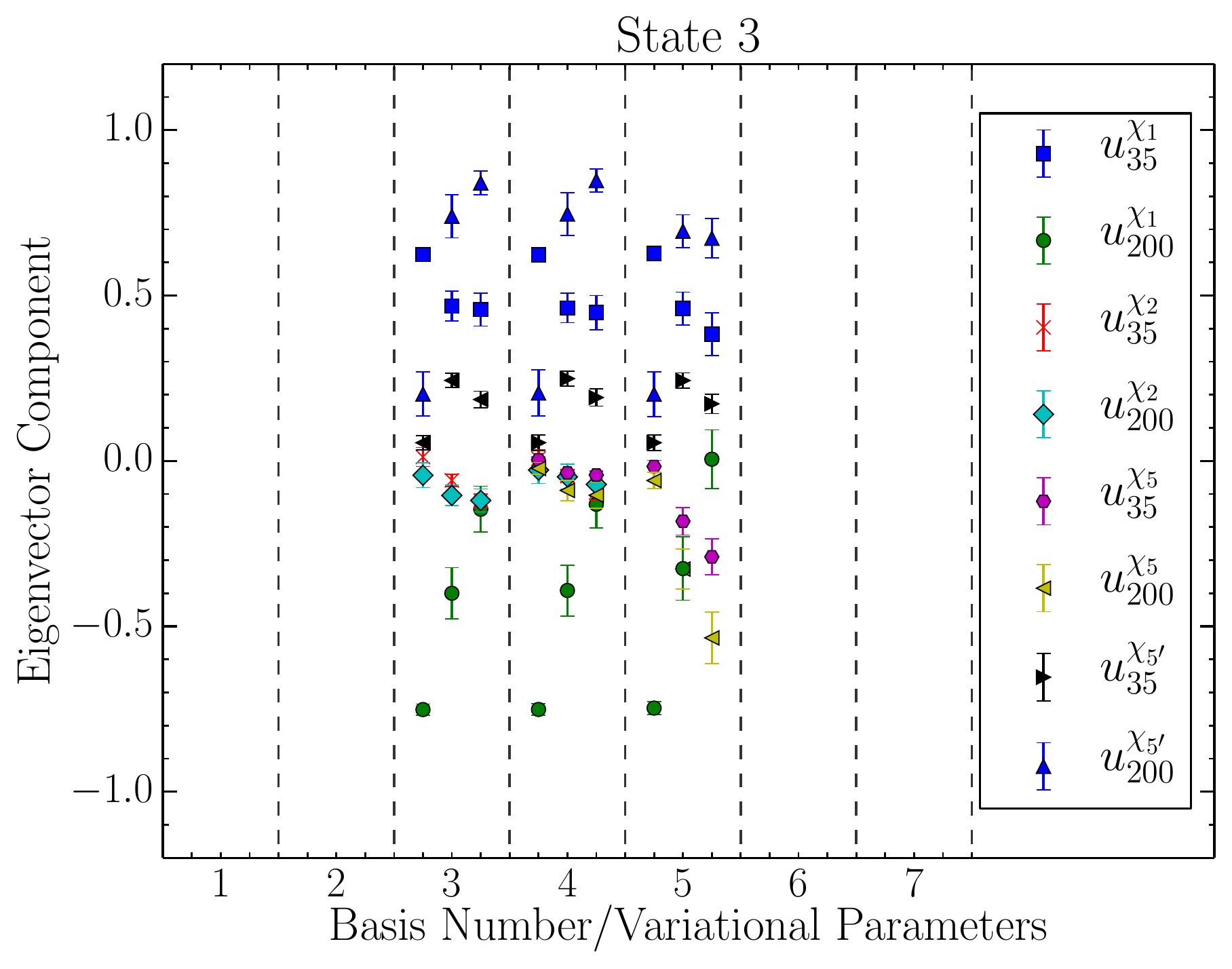}}
  \caption{(Colour online).  Eigenvector components corresponding to low-lying negative-parity nucleon states.  States 1 and 2 correspond to the two lowest-lying resonant states, while state 3 interestingly lies in the region of the P-wave scattering thresholds.  The column numbers denote basis number while the minor $x$ axis ticks correspond to the values of the variational parameter $\dt$ which runs from 1 through to 3.  $t_{0} = 17$ has been used throughout.  The subscripts 35 and 200 in the legend refer to the number of smearing sweeps applied.}
 \label{fig:Evectors-s123}
\end{figure}

Plots of the corresponding eigenvectors for the low-lying negative-parity 
states as a function of basis and variational parameter
$\dt=1\ldots 3$ are presented in Figures ~\ref{fig:Evectors-s0} and ~\ref{fig:Evectors-s123}. The upper
limit of $\dt=3$ was chosen as the largest value for which the
variational analysis converged for all seven bases.  The eigenvector
components for state 0 (when it is present) are dominated by the
multi-particle operators $\chi_5$ and $\chi_5',$ suggesting that this
state should be identified as a scattering state. The extracted energy
for this state is in the region of the non-interacting S-wave $N\pi$ scattering
threshold (which lies below the first negative-parity resonant
state). The uncertainty in bases 6 and 7 are relatively large compared
to basis 5, indicating that the presence of $\chi_1$ may also be
required to cleanly isolate this scattering state. Indeed, we note
that in basis 5 there is a significant contribution to state 0 from
the $\chi_1{(n_s=200)}$ operator.

It is also important to note that either $\chi_5$ or $\chi_5'$ can be the dominant interpolator exciting this lowest-lying state.  Given that $\chi_2$ is predominantly associated with the third state in the positive-parity sector at 2.4 GeV one might naively expect $\chi_5'$ would be associated with S-wave scattering states near 2.7 GeV.  Remarkably it creates a scattering state near 1.35 GeV.  Thus one should use caution in predicting the spectral overlap of five-quark operators by examining the spectral overlap of the pion and nucleon components of the five-quark operators separately.  In light of the quark field operator contractions required in calculating the full two-point function this result is not surprising.

In accord with previous studies~\cite{Mahbub:2012ri, Mahbub:2013ala},
we find that the $\chi_{1}$ interpolating field is crucial for
extracting state 1, associated with the lowest-lying negative-parity resonance, as we
do not observe this state when $\chi_{1}$ is absent as in bases 6 and
7. As expected, $\chi_{1}$ provides the dominant contribution to state
1, which is associated with the $S_{11}(1535)$ in Nature. Similarly,
we see that $\chi_{2}$ has a high overlap with state 2, the next
resonant state.  Basis 5 does not see state 2 due to the absence
of $\chi_{2}$.  However, unlike state 1, there is an important mixing of $\chi_1$ and $\chi_2$ in isolating the eigenstate.  It is interesting to note that in basis 7 we are able to form this state by combining $\chi_5$ and $\chi_5'.$

The consistency of the eigenvector structure for the low-lying states 1 and 2 is strong. Despite the appearance of a state near the
S-wave $N\pi$ threshold, state 0 in basis 5, the eigenvector components for
state 1 are remarkably consistent with those in other bases where this lower-lying state is absent.  If we look at basis 6, where state 0 is
present but state 1 is absent, the eigenvector components for state 2
are in good agreement with those from other bases where the
lower-lying state 0 is not observed.  This demonstrates that, with a
judiciously chosen variational technique, a reliable analysis of
higher states in the spectrum can be performed even if states associated with the
low-lying scattering states are not extracted by the correlation
matrix analysis.

\begin{figure}[htbp]
  \centering 
  {\includegraphics[width=0.48\textwidth]{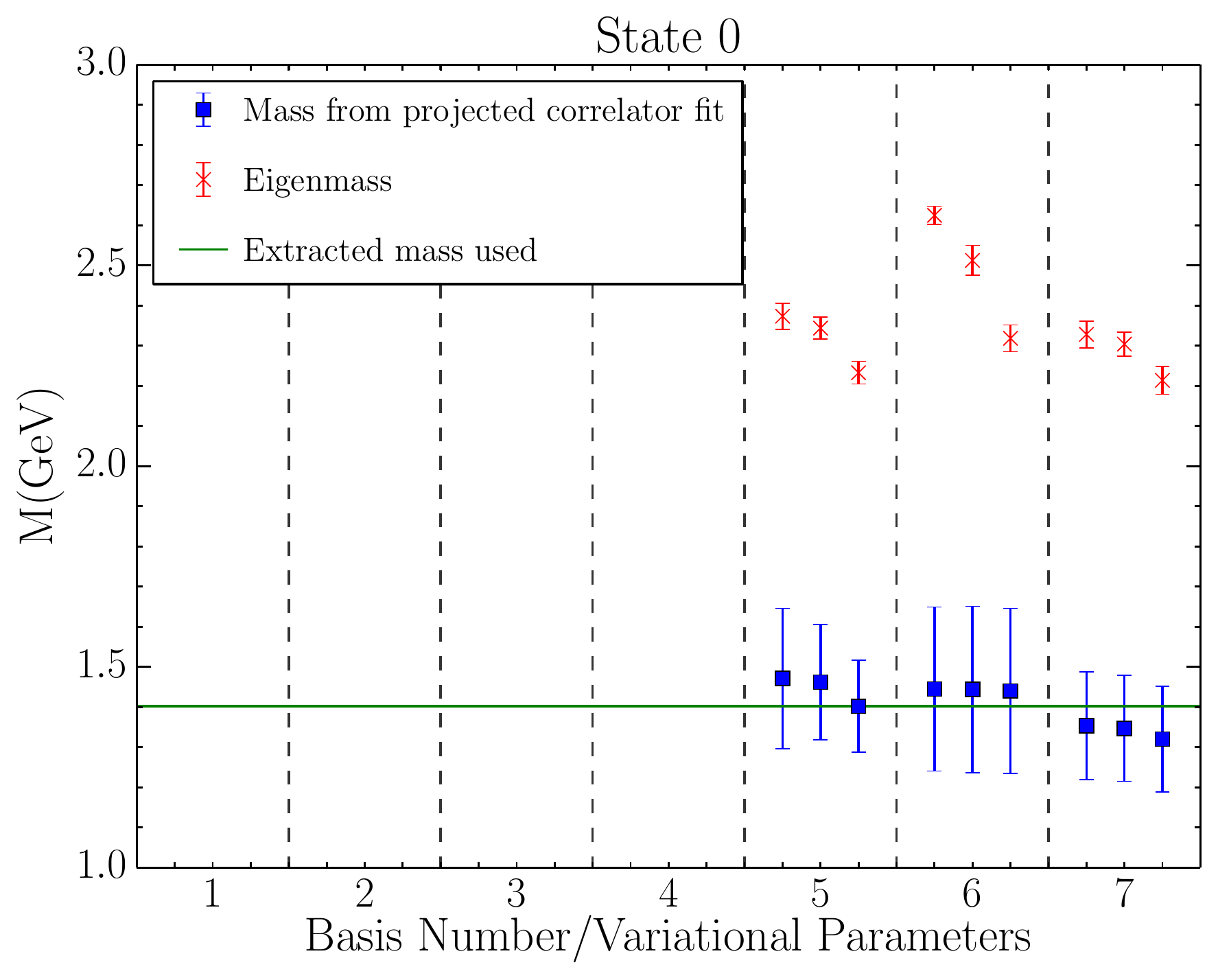}}
\caption{(Colour online).  Comparisons of eigenmasses to masses obtained from a projected correlator fit for state 0, which is in the region of the non-interacting S-wave $N\pi$ scattering threshold.  The column numbers denote basis number while the minor $x$ axis ticks correspond to the values of the variational parameter $\dt=1\ldots 3.$  $t_{0} = 17$ has been used throughout.  The line denoting the extracted mass used has been set using basis 5 with $\dt=3.$}
\label{fig:EmassvsProjMass-s0}
\end{figure}
\begin{figure}[htbp]
  \centering 
  {\includegraphics[width=0.48\textwidth]{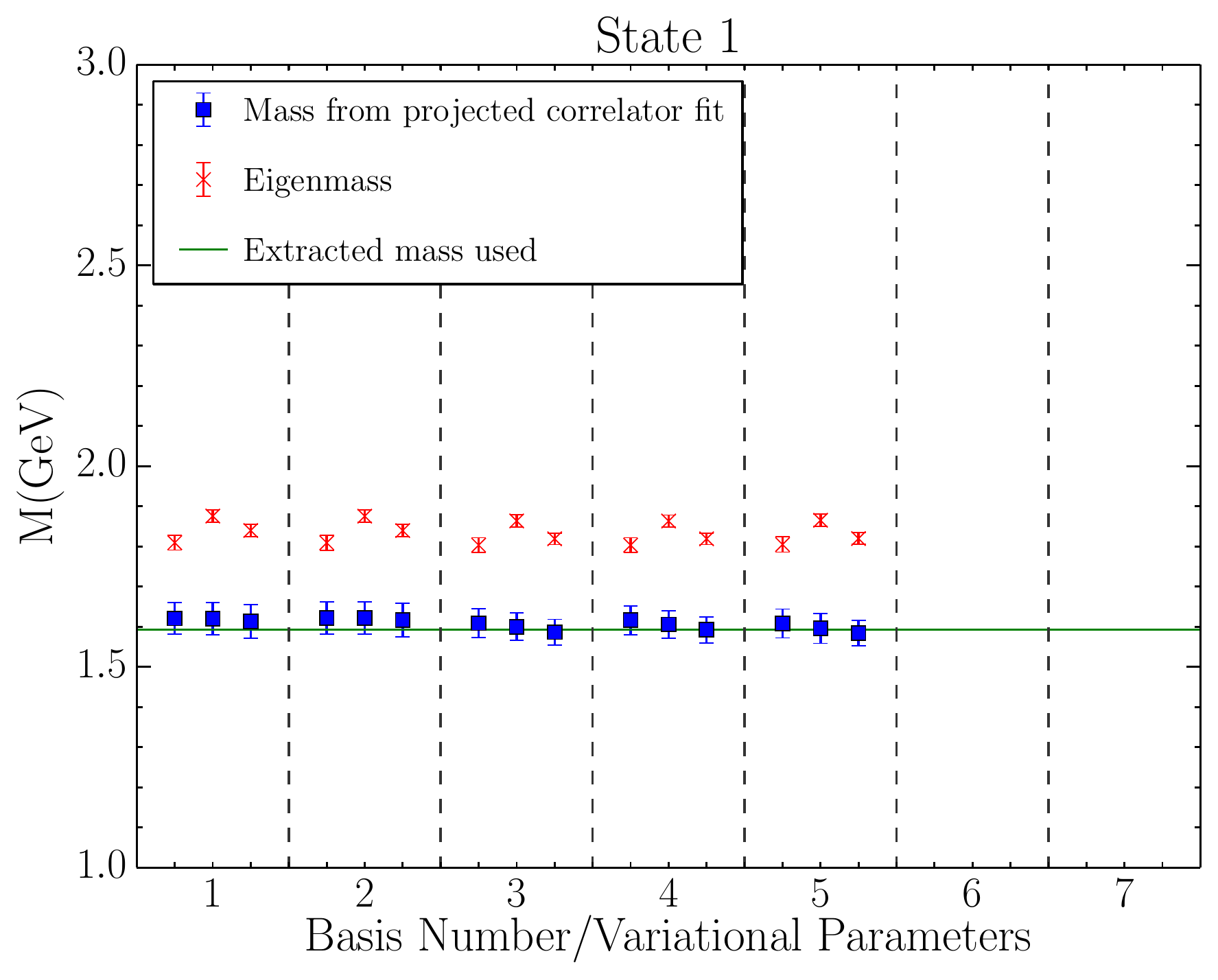}}\\
  {\includegraphics[width=0.48\textwidth]{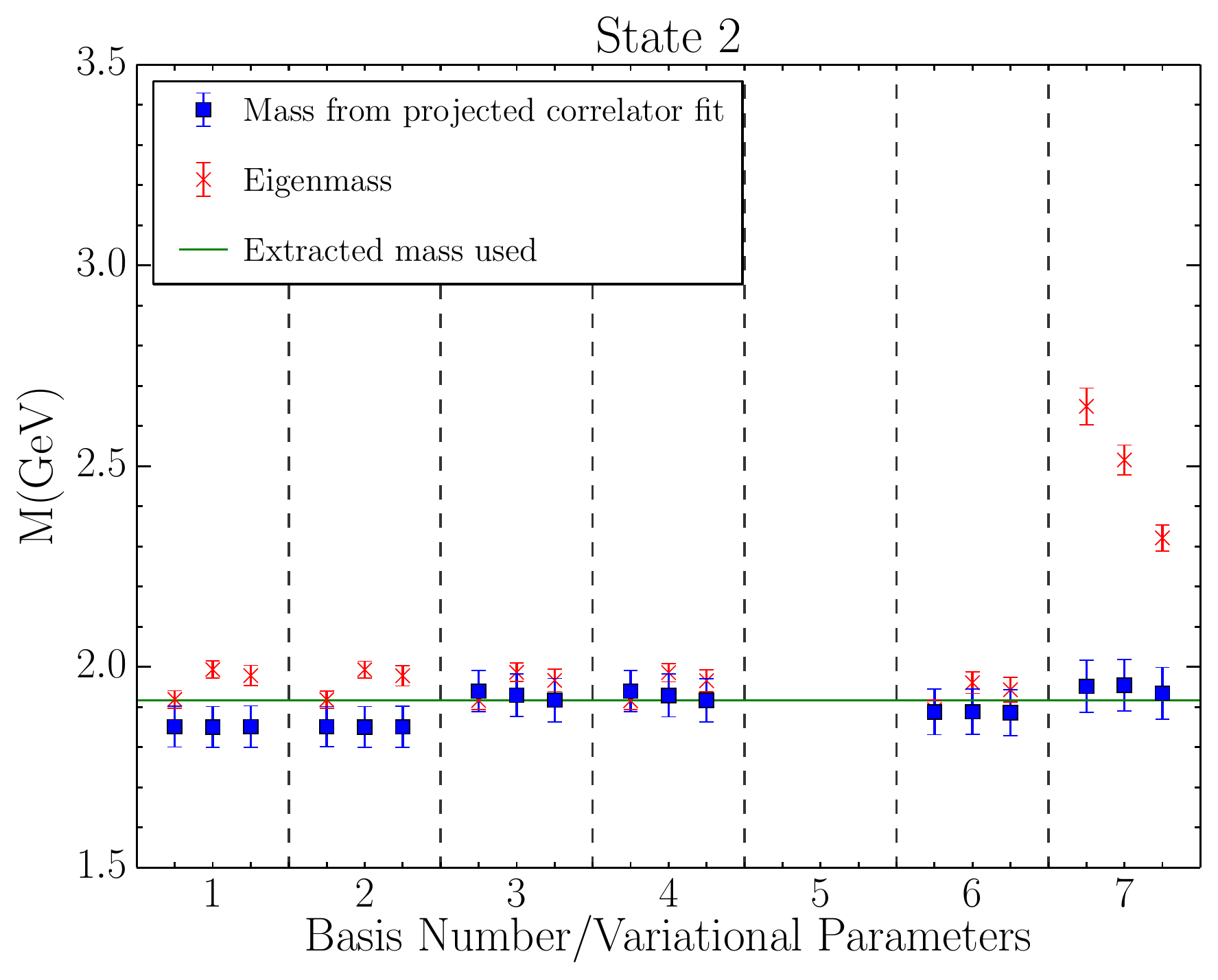}}\\
  {\includegraphics[width=0.48\textwidth]{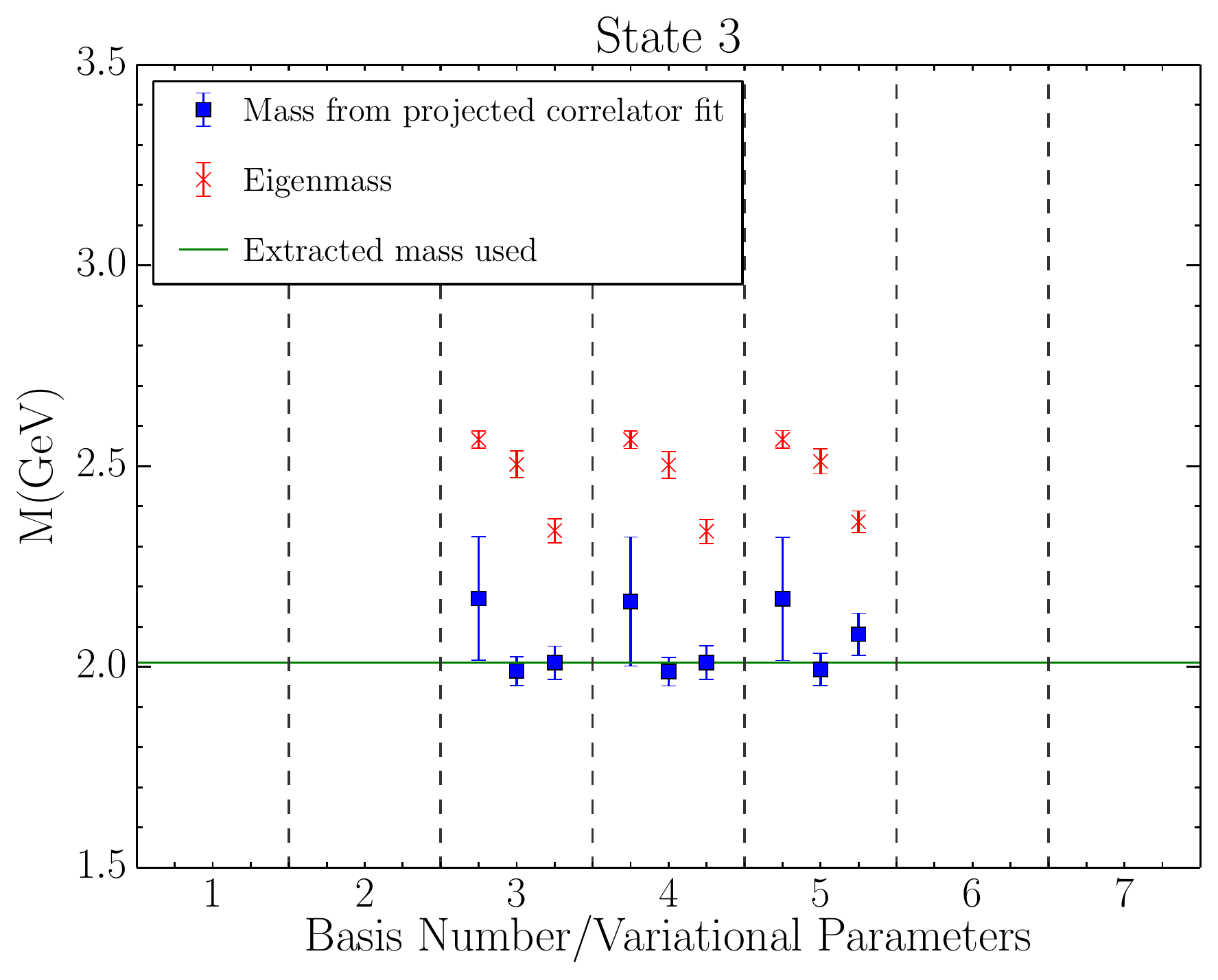}}
\caption{(Colour online).  Comparisons of eigenmasses to masses obtained from a projected correlator fit for low lying states in the negative-parity nucleon channel. The column numbers denote basis number while the minor $x$ axis ticks correspond to the values of the variational parameter $\dt=1\ldots 3.$  $t_{0} = 17$ has been used throughout.  The line denoting the extracted mass used has been set using basis 4 with $\dt=3.$}
\label{fig:EmassvsProjMass-s123}
\end{figure}

State 3, which lies in the region of the non-interacting P-wave $N\pi\pi$
scattering states in the channel, also shows good agreement across bases and
variational parameters.  The dominant eigenvector components show that
this state is formed from a mix of $\chi_5'$ and $\chi_1.$ It is worth
noting that very early choices of the variational parameters such as
$(t_{0},\dt) = (17,1)$ do not allow sufficient Euclidean time evolution
to cleanly isolate this state. The correlation matrix has more states
participating in the analysis than the dimension of the basis leading
to contamination from unwanted states and hence spurious results.  The
different structure for the state 3 eigenvectors at these early
variational parameter sets illustrates the need to allow sufficient
Euclidean time evolution to occur.

The comparison of the fitted masses as a function of variational
parameter $\dt$ across the different bases for the negative-parity
sector is shown in Figures \ref{fig:EmassvsProjMass-s0} and \ref{fig:EmassvsProjMass-s123}. Again, the
eigenmasses are plotted for comparison. As before, we observe for all
the states the fitted masses are consistent across the different bases
and values of $\dt$.  In contrast, the eigenmasses for the negative-parity states all show
some variation with $\dt$ to different extents, with the values typically lying well
above the extracted energies.

Finally, we observe that whenever $\chi_{5}^{\prime}$ is present, either a
state near the S-wave $N\pi$ scattering threshold, or a state lying in the region of the P-wave $N\pi\pi$ scattering thresholds is extracted.  This indicates the presence of the
vector di-quark in the interpolator may play an important role in
scattering state excitation.  It is perhaps surprising that
basis 4 fails to see a state near the lowest-lying scattering threshold in the
sector, despite being the largest basis.  We believe this is due to the spectral strength available to the scattering state being relatively low.
The overlap of the scattering state with the operators is not 
high enough to compete with the
large spectral strength imparted to the low-lying resonant states when both
$\chi_1$ and $\chi_2$ are present.  We note that the only time our
local (three-quark or five-quark) operators overlap with a meson-baryon state is when both hadrons are at the origin. The
probability of this occurring is proportional to $1/V^2.$ After taking
into account the spatial sum in Eq. \ref{defn:CM}, this results
in a $1/V$ suppression of multi-particle states in the correlator
amplitude $\mathcal{G}(t)$~\cite{Luscher:1991cf}. Indeed, it seems
to be relatively difficult to extract a state near the S-wave $N\pi$ state with our
local five-quark operators, suggesting that scattering state
excitation is best achieved by explicitly projecting the momentum of
interest onto each hadron present in the scattering state.

%
%

\section{Conclusions}
\label{sect:Conclusions}

We have investigated the role of local multi-particle interpolators in
calculating the nucleon spectrum by examining a variety of different
bases both with and without five-quark operators. 

The variational
techniques herein employed, demonstrate that fitting a
single-state ansatz to optimised eigenstate-projected correlators
provides a method to reliably extract energies in both the positive
and negative-parity channels.  While the selection of states that are
observed varied between bases, when a given state is seen the
extracted energy agrees across qualitatively different bases.

Furthermore, the structure of the eigenvector components and the
corresponding fitted energies for the states observed are shown to be
highly consistent across different bases and choices of the
variational parameters, despite the markedly different interpolators
used in the various bases. We found that an approximate accidental degeneracy in
the eigenmass at $(t_0,\dt) = (17,2)$ for states 2 and 3 led to a
large increase in the uncertainties for the corresponding energies and
eigenvector components.

While we did not observe any positive-parity scattering states, in the negative-parity sector we found that $\chi_{5}^{\prime}$ was crucial to obtaining an energy in the region of the non-interacting S-wave $N\pi$.  Even with
the use of local five-quark interpolators the uncertainties on this 
threshold state were relatively large compared to those
of higher states.  Future studies will include multi-particle operators
with explicitly projected single-hadron momenta in the variational
basis to facilitate better excitation of scattering states, including
those in the positive-parity sector.
An interesting feature of our negative-parity results is that the
energies of the extracted states are consistent across all
bases in which the state is observed, regardless of the presence (or
not) of a state in the region of the lower-lying non-interacting scattering threshold.  This suggests that by using the techniques
described herein, one does not need to have access to the aforementioned low-lying states
to reliably extract energies closely related to the resonances of Nature.

\section*{Acknowledgments}
We thank Mike Peardon for helpful discussions of the stochastic noise technology.  
We thank the PACS-CS Collaboration for making these $2+1$ flavor
configurations available and the ongoing support of the ILDG.  This
research was undertaken with the assistance of resources at the NCI
National Facility in Canberra, Australia, and the iVEC facilities at 
the University of Western Australia (iVEC@UWA). 
These resources were provided through the
National Computational Merit Allocation Scheme, supported by the
Australian Government and the University of Adelaide Partner Share.
We also acknowledge eResearch SA for their supercomputing support
which has enabled this project.  This research is supported by the
Australian Research Council.

\bibliographystyle{apsrev4-1}
\bibliography{reference}

\end{document}